\documentclass[conference]{IEEEtran}
\IEEEoverridecommandlockouts
\usepackage{amsmath,amssymb,amsfonts}
\usepackage{algorithmic}
\usepackage{graphicx}
\usepackage{textcomp}
\usepackage{xcolor}
\def\BibTeX{{\rm B\kern-.05em{\sc i\kern-.025em b}\kern-.08em
   T\kern-.1667em\lower.7ex\hbox{E}\kern-.125emX}}
   
\usepackage{booktabs} 
\usepackage{array}
\usepackage{mathtools}
\usepackage{url}
\usepackage{listings}
\usepackage{balance}
\usepackage{tikz,color,subfigure}
\usepackage{enumitem}
\usepackage{courier}

\usepackage{caption}
\usepackage{flushend}

\usepackage{color,soul}

\usepackage[numbers]{natbib}
\usepackage{microtype}

\newcommand{\abbr}{SimFS}

\setitemize{topsep=1pt,parsep=0pt,partopsep=0pt,leftmargin=1.5em}

\makeatletter
\newdimen\SOUL@dimen 
\def\SOUL@ulunderline#1{{%
    \setbox\z@\hbox{#1}%
    \SOUL@dimen=\wd\z@ 
    \dimen@i=\SOUL@uloverlap
    \advance\SOUL@dimen2\dimen@i 
    \rlap{%
        \null
        \kern-\dimen@i
        \SOUL@ulcolor{\SOUL@ulleaders\hskip\SOUL@dimen}
    }%
    \unhcopy\z@
}}

\makeatother

\begin{document}

\author{\IEEEauthorblockN{Salvatore Di Girolamo}
\IEEEauthorblockA{\textit{Dept. of Computer Science} \\
\textit{ETH Zurich}\\
Switzerland \\
digirols@inf.ethz.ch}
\and
\IEEEauthorblockN{Pirmin Schmid}
\IEEEauthorblockA{\textit{Dept. of Computer Science} \\
\textit{ETH Zurich}\\
Switzerland \\
schmidpi@student.ethz.ch}
\and
\IEEEauthorblockN{Thomas Schulthess}
\IEEEauthorblockA{\textit{Dept. of Physics} \\
\textit{ETH Zurich}\\
Switzerland \\
schulthess@cscs.ch}
\and
\IEEEauthorblockN{Torsten Hoefler}
\IEEEauthorblockA{\textit{Dept. of Computer Science} \\
\textit{ETH Zurich}\\
Switzerland \\
htor@inf.ethz.ch}
}

\makeatletter
\let\@@pmod\pmod
\DeclareRobustCommand{\pmod}{\@ifstar\@pmods\@@pmod}
\def\@pmods#1{\mkern4mu{\operator@font mod}\mkern 6mu#1}
\makeatother

\title{SimFS: A Simulation Data Virtualizing \\ File System Interface}
\maketitle

\begin{abstract}
Nowadays simulations can produce petabytes of data to be stored in parallel
filesystems or large-scale databases.  This data is accessed over the course of
decades often by thousands of analysts and scientists. However, storing these
volumes of data for long periods of time is not cost effective and, in
some cases, practically impossible. 
We propose to transparently virtualize the simulation data, relaxing the
storage requirements by not storing the full output and re-simulating the
missing data on demand. We develop SimFS, a file system interface that exposes
a virtualized view of the simulation output to the analysis applications and
manages the re-simulations.  SimFS monitors the access patterns of the analysis
applications in order to (1) decide the data to keep stored for faster accesses
and (2) to employ prefetching strategies to reduce the access time of missing
data. 
Virtualizing simulation data allows us to trade storage for computation: 
this paradigm becomes similar to traditional on-disk analysis 
(all data is stored) or in situ (no data is stored) according with
the storage resources that are assigned to SimFS. 
Overall, by exploiting the growing computing power and relaxing the storage 
capacity requirements, SimFS offers a viable path towards exa-scale simulations.
\end{abstract}

\section{Motivation}


Reliable long-term data archiving is very costly. For example, storing
10 TiB for 10 years costs between \$2,400 and \$6,000 on Microsoft's
Azure. 
The only practical scheme to mitigate these costs, besides deletion, is
(lossy or lossless) compression of the data and it is fundamentally
constrained by the tradeoff between data size and quality. 
%
When taking a closer look at \emph{how data is generated}, we observe
two fundamentally different modes: (1) data collected by sensors or
terminals that observe non-deterministic environments or (2) data
generated by deterministic simulations that model complex
and potentially chaotic systems. 
\emph{We observe, that the latter could be recomputed on demand instead
of stored, given the right data retrieval system.}

Many simulation applications produce vast amounts of data that is today
stored in large filesystems or databases.
For example, the European Centre for Medium-Range Weather Forecasts
(ECMWF) alone had an archive of 100 PiB in 2015, experiencing an annual
growth rate of 45\%~\cite{Grawinkel:2015:AES:2750482.2750484}; 
by 2020, their archive will reach a Zettabyte.
Climate model data is used by countries and insurances to make critical
decisions thus repeatability of analyses is mandated by international
regulatory bodies. 
Astrophysics simulations are another example where data volumes grow
with the compute capabilities, creating more than 20 PiB of
data each~\cite{potter2016pkdgrav3}. Thousands of such simulations are
collected in virtual observatories, mainly limited by the storage
costs~\cite{bernyk2016theoretical,otv}. 
Those two examples outline a clear trend: As we proceed into the age of
simulation~\cite{winsberg2010science}, \emph{big (simulation) data} will
soon be required for many real-world decisions.

The data produced by large simulations is commonly used by thousands
of analysts and scientists over the course of decades.  They are used in
analysis workflows where the data is stored in files or databases.
Specifically, these workflows address two requirements: (1) data can
conveniently be analyzed with any access pattern (e.g., time-reverse
or random access) and (2) the exact same data can (often years) later be
re-analyzed to reproduce the results. 
%
%
This makes the data-backed analysis a de-facto standard for today's
simulation data analytics. 

\begin{figure}[h]
    \vspace{-1em}
    \centering{
        \includegraphics[trim=0 0 0 0, clip, width=1\columnwidth]{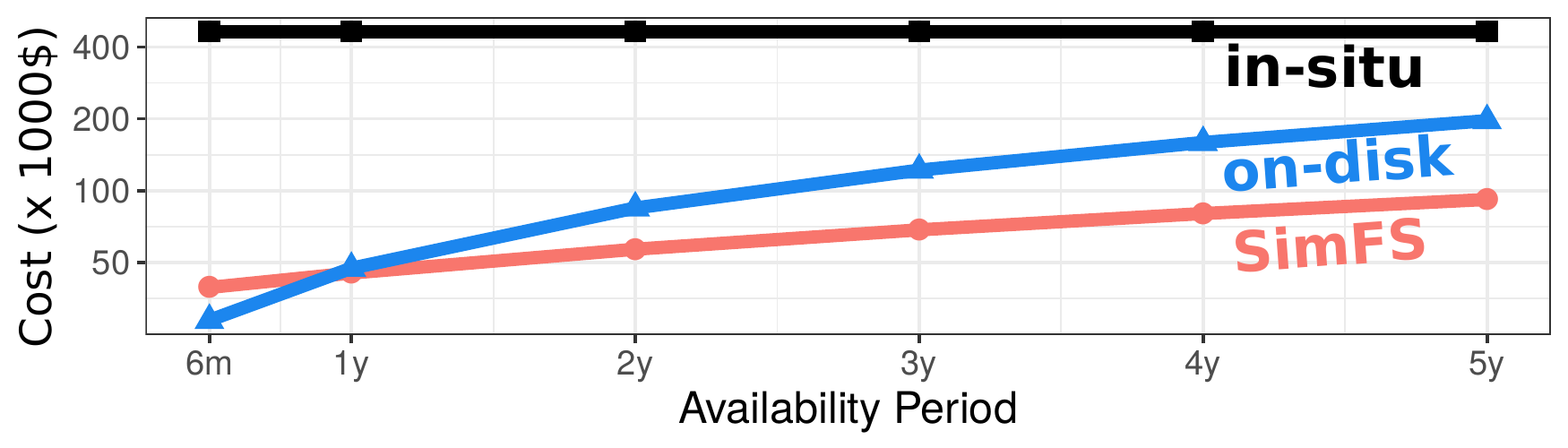}
    }
    \vspace{-1.2em}
    \caption{Aggregated analysis cost. The cost 
    of the different analysis solutions (on-disk, in-situ, \abbr{})  
    is function of the time period over which the analyses are executed.}
    \label{fig:cost_teaser}
\end{figure}

We propose \emph{SimFS}, a file system interface that virtualizes simulation
output data for analysis tools. SimFS avoids storing the whole
simulation output data but stores checkpoints to re-start parts of the
simulation and produce missing files \emph{on demand}. A virtualized
view, similar to virtual memory, is provided to the analysis tools, enabling
them to work \emph{as if} all output data exists as files. This way,
SimFS can exploit the tradeoff between inflexible
\emph{in-situ} analysis, where all analyses are running together with
the simulation and no data is stored, and \emph{on-disk}, where
the full simulation output is stored and no re-simulations are needed.
Figure~\ref{fig:cost_teaser} shows the expected costs for performing 100
analyses equally spaced over varying data availability periods for a
real-world climate simulation scenario discussed in detail in
Sec.~\ref{sec:costeff}. It shows that SimFS can reduce the costs for
a five-year period from more than \$200,000 for an on-disk solution to
less than \$100,000. We also show ``in-situ'', which re-runs the whole
simulation for each analysis as comparison. 


Implementing \abbr{} poses interesting challenges that we describe in the
following. 
In the past, computation speeds and cost efficiency grew much faster
than storage speeds and efficiency. Whether this trend continues or not,
\abbr{} must always adjust to the exact cost and performance tradeoff. 
While the file-system virtualization itself is simple, SimFS
employs complex caching and prefetching strategies to adjust the
tradeoff between computation (resimulation) and storage cost.
To guide optimizations, it exposes a set of interfaces that can be used
\emph{in addition} to the fully transparent virtualization to optimize
client applications as, e.g., guided prefetching or non-blocking
reads. 
By nature of the virtualization, SimFS transparently enables large-scale
analyses on multi-petabyte datasets on terabyte storage systems that
have been impossible so far.
Thus, SimFS not only enables new scientific breakthroughs but it also
allows the system cost to shrink with the computation costs.

\begin{figure}[h]
\centering{
    \vspace{-1.1em}
    \includegraphics[trim=0 150 0 0, clip, width=1\columnwidth]{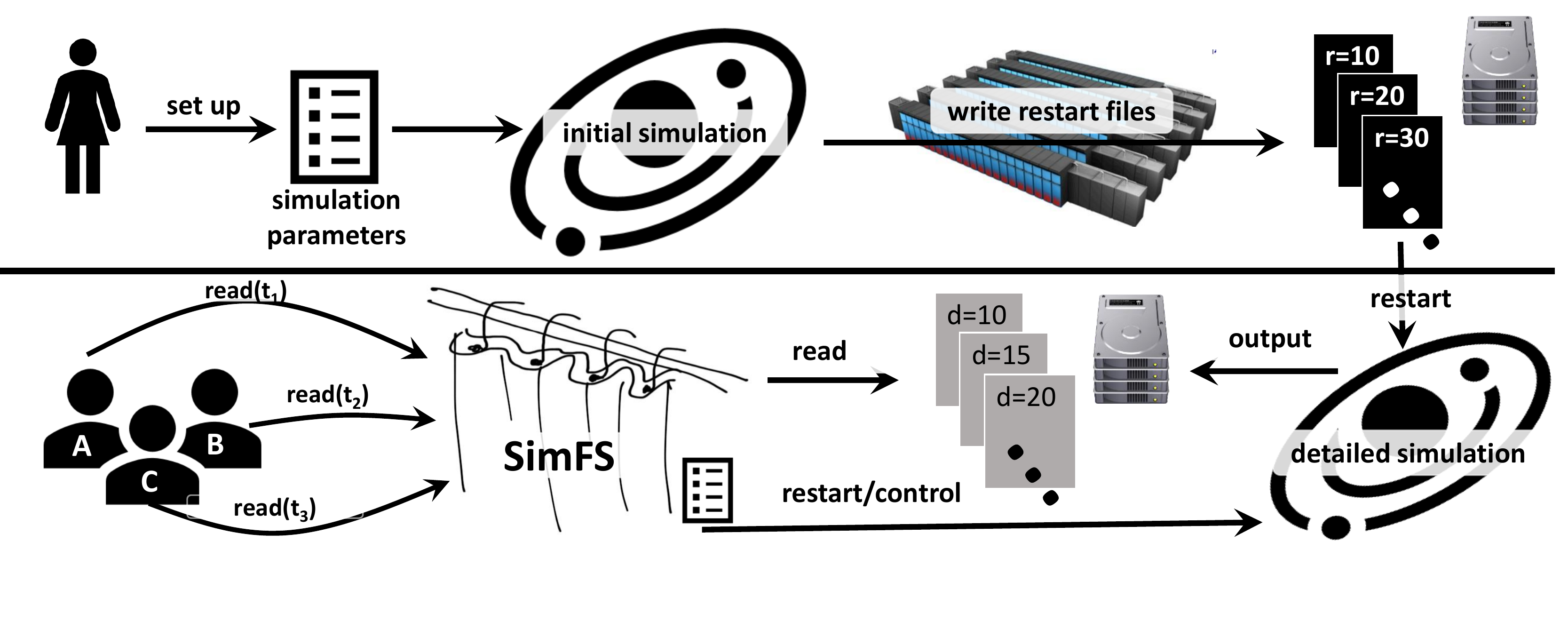}}
    \vspace{-3em}
    \caption{Overview of \abbr{}}
    \label{fig:overview}
    \vspace{-0.4em}
\end{figure}
Figure~\ref{fig:overview} shows the abstract workflow of SimFS:
The simulation is initially set up by a scientist (top left of
Figure~\ref{fig:overview}) and runs to completion while producing restart files
(black files, top right).  First in-situ analyses may be performed during the
initial simulation but we focus on later analyses. 
Later, analysis tools from different clients (e.g., researchers in the lower
left) access the virtualization layer through standard data-access interfaces
such as HDF5~\cite{folk1999hdf5}, netCDF~\cite{rew1990unidata}, or
ADIOS~\cite{lofstead2008flexible}.
\abbr{} manages the simulations to re-create output data (gray files, bottom)
on demand and delivers it to the analysis tools. We remark that simulations can
be restarted on different devices than the original simulation, e.g., smaller
GPU systems, because the simulated time intervals are less demanding. 

SimFS requires that the simulation can be re-started from checkpoints and
delivers a bitwise-identical output to the original run. While
checkpoint/restart facilities are already needed to deal with limited compute
time and failures, bitwise reproducibility may not generally be available.
However, it should generally be used for good scientific practice
(repeatability) and can be achieved with a set of standard techniques without
significant performance penalty~\cite{arteaga-bitrep, muller2018reproducible}.
If bitwise reproducibility cannot be guaranteed, we expect the analyses being
able to operate on data that is different from the one produced by the initial 
simulation. The analysis can check if the re-simulated data differs by using
the \abbr{} APIs.
%

We argue that \abbr{} solves a significant part of the big data storage
challenge in simulation sciences. We will show how it even improves
analysis performance and automatically utilizes available storage resources
efficiently, all without requiring any modifications of the analysis
tools. \abbr{} is used on some of the largest machines existing today.

%
%

%

\section{Virtualizing Simulations}\label{sec:simfs}

Virtualizing simulation data is very similar to virtual memory and paging, a
key component in today's operating systems: the simulation output is our
virtual memory and the pages are sets of simulation output files. If an
application accesses a file that is not on disk (i.e., swapped out), then the
entire page containing that file has to be re-simulated. (i.e., loaded).  While
memory is virtualized by means of memory loads and stores, we virtualize
simulation data by intercepting calls to I/O libraries.
In this model, \abbr{} acts like a memory management unit but on a coarser
grain. 
\begin{figure}[h]
    \vspace{-0.5em}
    \centering{
        \includegraphics[width=1\columnwidth]{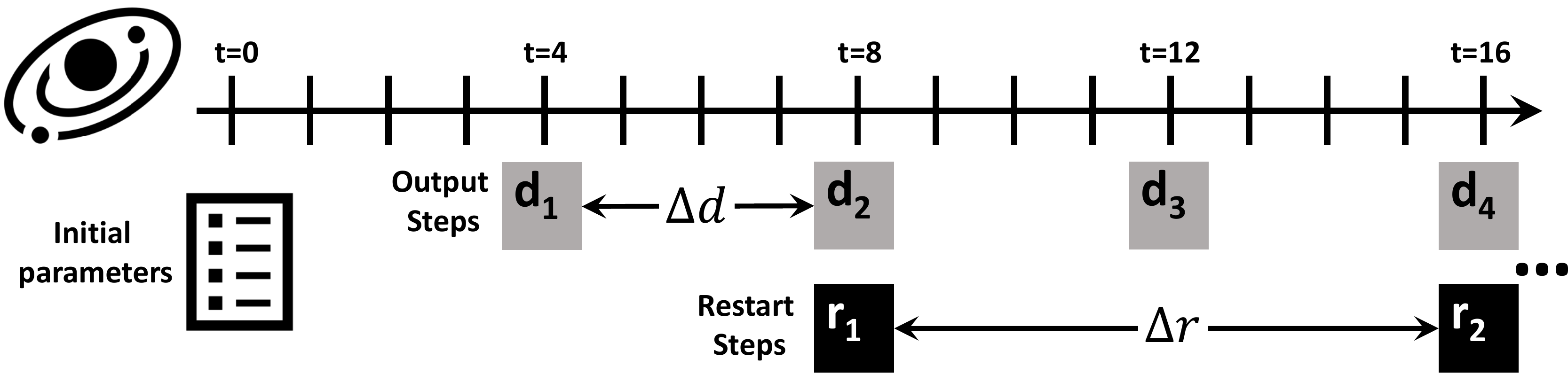}}
    \caption{Simulate time, output, and restart steps.}
    \label{fig:steps}
    \vspace{-1em}
\end{figure}

\subsection{Modeling Simulations}\label{sec:virmod}
Our model focuses on forward-in-time simulations. They generate one or more
\emph{output steps} during the run, each of which contains one or more
\emph{timesteps}. 
A single time step can encapsulate multiple smaller simulation steps that 
are not visible in the output steps, hence not exposed in our model.
Furthermore, simulators commonly provide the ability to
write restart steps that can be used to restart a simulation. Output and
restart steps are stored in files.

As simulations proceed in timesteps $t_1$, $t_2$, $\ldots$, $t_n$, a simulator
configuration is defined by $\Delta d$, that is the number of timesteps between
two output steps, and $\Delta r$, that is the number of timesteps between two
restart steps. Each output step contains all $\Delta d-1$ timesteps beginning
from either the last output step or the beginning of the simulation. We assume
that the simulation can be restarted from any restart step and proceeds forward
in time. Thus, to produce an output step $d_i$, the simulation needs to be
started from the closest previous restart step \mbox{$R(d_i) = \lfloor \frac{i
\cdot \Delta d}{\Delta r}\rfloor$}.  To exploit spatial locality, we let a
re-simulation run until at least the next restart step 
\mbox{$\lceil \frac{i \cdot \Delta d}{\Delta r}\rceil$}.
Choosing $\Delta d$ and $\Delta r$ allows us to adjust the
time-space tradeoff. If \abbr{} stores all output steps, we can serve all
requests from the output files directly. However, we assume that we cannot
store the complete output on disk. Then, $\Delta d$ selects the granularity of
the data generation and $\Delta r$ the time to reach a specific timestep. In
particular, the bigger the $\Delta r$, the lower the number of restart files
that need to be stored and the higher the average time to simulate a specific
output step.

Figure~\ref{fig:steps} shows an example where a simulation starts from
$t=0$ and runs forward-in-time beyond $t=16$ (not shown
in the figure).  Each output step contains four timesteps ($\Delta d=4$) and
can be restarted every 8 timesteps ($\Delta r=8$).

\textbf{Simulation Contexts}
Simulation output characteristics (i.e., output steps content, $\Delta d$,
$\Delta r$) are determined by a specific simulation configuration and a
simulator can have multiple configurations.
We define a simulation context as a simulator and an its configuration.  Since
the analysis applications operate on the simulation output produced by a given
context, simulation contexts are a central component of our model.

Multiple simulation contexts can share the same restart files, offering
different simulation outputs that can be produced at different speeds.
Analyses can be interested in one or more simulation output types, hence in one
or more simulation contexts: e.g., analyzing a coarser grain simulation output
on a simulation context and then switch to finer grain on a different context
for a more detailed study of interesting events.

For a given simulation, scientists identify multiple simulation contexts that
are made available to the analyses through \abbr{}. Since each simulation
context can produce different subset of output steps, they can lead to
different re-simulations costs.  The analyses can specify their simulation
context via an environment variables or the \abbr{} APIs.

\begin{figure}[b]
    \vspace{-1.7em}
    \centering{
        \includegraphics[trim=20 18 20 10, clip, width=1\columnwidth]{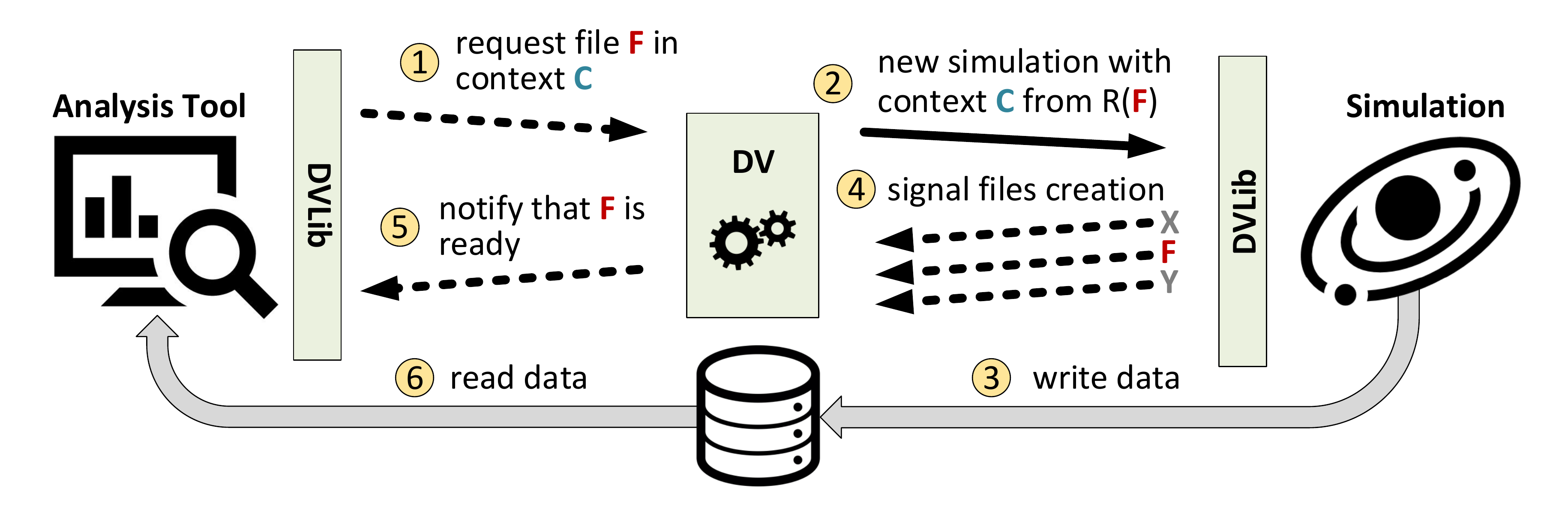}}
    \vspace{-1.6em}
    \caption{Handling misses in \abbr{}. Dashed arrows are control messages
             (TCP/IP); solid (thin) arrows represent actions (script execution); 
             solid (bold) arrows represent data movement (file system).}
    \label{fig:miss}
    \vspace{-1em}
\end{figure}

\section{SimFS}

\abbr{} consists of two components: (1) the \emph{Data Virtualizer} (DV), a
daemon process that coordinates simulations and analyses and (2)
the \emph{DV Library} (DVLib), that enables the analyses and
simulations to communicate and synchronize with the DV.
DVLib provides bindings for many I/O libraries (e.g., netCDF, HDF-5) so the
analyses and simulations can be transparently interfaced to the DV. Moreover,
it exposes a set of APIs to let virtualization-aware analysis applications
have a more direct control on the virtualized environment. 

\subsection{Virtualized Simulation Output Analysis}\label{sec:simfs_ex}
Analysis accesses to the simulation output are intercepted by DVLib, which
communicates with the DV to check if the requested files are available. 
%
After intercepting an \textbf{open} call issued by an analysis application, the
DVLib sends a request to the DV and waits for a response. If the file exists,
then an acknowledgment is sent back to the application, which is now free to
open the file. 
Figure~\ref{fig:miss} shows the case in which the requested file is not
available. 
\includegraphics[trim=15 19 375 18, clip, width=0.15in]{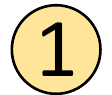}
Once DV receives the request and checks that the file is not available, 
\includegraphics[trim=90 19 300 18, clip, width=0.15in]{visio/numbers.pdf}
it starts a new re-simulation, configured according
with the context specified by the analysis application.
\includegraphics[trim=160 19 228 18, clip, width=0.15in]{visio/numbers.pdf}
The new simulation starts producing output steps, writing them on the (parallel)
file system. 
\includegraphics[trim=230 19 160 18, clip, width=0.15in]{visio/numbers.pdf}
DVLib is aware of the files created by the running simulation since it
intercepts the \textbf{close} calls issued by it.  Once a file is closed, DVLib
assumes that this file is ready on disk and notifies the DV of this event. 
\includegraphics[trim=300 19 90 18, clip, width=0.15in]{visio/numbers.pdf}
When notified, DV checks if there are analysis applications waiting for the
file and, if yes, it forwards the notification to them. 
\includegraphics[trim=375 19 18 19, clip, width=0.15in]{visio/numbers.pdf}
After receiving the notification from DV, the DVLib running on the analysis
call unblocks and perform the original I/O library call that will now find the
file on disk.

A key motivation of virtualizing simulations output is to enable the analysis
of datasets that are one or more orders of magnitude larger than the available
storage capacity. This implies that \abbr{} has to monitor the data volume
occupied by a simulation context and eventually evict output steps when the
given storage resources are saturated. 
In particular, we associate each simulation context with a storage area 
(i.e., a file system directory). When a new re-simulation from a given context
is launched, DVLib intercepts the \textbf{create} calls from the simulator and
redirects them to the associated storage area.
The simulation context also specifies the maximum size of its storage area.
When the actual size of a storage area reaches its maximum size, the DV applies
eviction policies (see Sec.~\ref{sec:caching}) for selecting one or more output
steps to evict.
\abbr{} associates a reference counter to each output step to keep track of
the analysis that are currently accessing it. An output step can
be evicted only if its reference counter is zero.

\enlargethispage{\baselineskip}

\subsection{Simulator Interface}

The DV is in charge to restart simulations to produce data that is being
accessed by the analysis applications but not on disk. 
However, how to configure and start a simulation is strictly related to the
simulator and to the system where the simulation has to be run  (e.g., how to
set start and stop output steps; how to submit the simulation job to the batch
system).  To let a simulator be managed by \abbr{} we introduce a
\emph{simulation driver} that can be implemented as a LUA script and provides
the following simulator-specific functionalities: 
\begin{itemize}[noitemsep,topsep=0pt,parsep=0pt,partopsep=0pt,leftmargin=*]
\item \textbf{Naming Convention:} The output steps file names follow a convention
that is specified by the simulator and its configuration. 
\abbr{} needs to be able to compare filenames for, e.g., finding the closest in
time restart step from which the simulation can be restarted to produce a
missing file. 
The simulation driver provides a function $key$ that given a filename, returns
a integer key such that if the output step $d_i$ is produced after $d_j$ by the
simulator, then $key(d_i) > key(d_j)$.

\item \textbf{Simulation Job}: When creating a new simulation, \abbr{} invokes
a simulation driver function that takes as arguments the simulation start and
stop output steps keys and the \emph{parallelism level}. This function creates
a script that the DV can execute to start the new simulation. \abbr{} needs to
tune the simulation parallelism to enable the optimizations described in
Sec.~\ref{sec:prefetching}. However, the simulator can impose constraints on
its resources allocation (e.g., square or power of two number of
processes). By using the parallelism level parameter, that is an integer from
$0$ to \emph{max parallelism level} (i.e., a parameter set by the simulation
driver), \abbr{} can increase the simulation parallelism without having to
directly enforce these constraints, which are instead enforced by the
simulator-specific implementation of the simulation interface.  
\end{itemize}
We intercept the \textbf{create} and \textbf{close} calls
issued by the simulator to let the DV trigger replacement policies and analysis
notifications, respectively. The mapping of these calls to standard I/O libraries
is reported in Table~\ref{table:opmap}.

%

\subsection{Analysis Application Interface}\label{sec:apis}

The analysis applications are interfaced to the DV through DVLib.  DVLib
provides bindings to standard I/O libraries, allowing legacy analysis
applications to transparently access virtualized simulation output, and a set of
APIs that can be used by virtualization-aware analysis applications.

\subsubsection{Transparent Mode}\label{sec:interface_transparent}
DVLib provides mappings to standard I/O libraries to enable analysis of
virtualized simulation output without requiring code changes to the analysis
applications or simulators. 
This is achieved by intercepting the \emph{open}, \emph{create}, \emph{read},
and \emph{close} calls of the different I/O libraries. Table~\ref{table:opmap}
shows the function names of these calls for the different I/O libraries we
provide bindings for.  When DVLib intercepts an \textbf{open} call, it sends a
request to the DV that checks whether the file exists. If not, a new simulation
is restarted as in Sec.~\ref{sec:simfs_ex}. This call is non-blocking, even if
the opened file is not on disk. When the applications tries to \textbf{read}
from a file that is not on disk, DVLib blocks the call (by not returning from
it) until the DV sends a notification of the file being ready. The
\textbf{close} call is intercepted to let DV decrease the
output step.
The simulation context name accessed by an analysis transparently interfaced
to \abbr{} can be specified as an environment variable.
\setlength{\tabcolsep}{0.2em}
\begin{table}[t]
    \vspace{-0.2em}
    \scriptsize
    \center
    \begin{tabular}{lllll}
        \toprule
        \textbf{Call} & \textbf{(P)NetCDF} \cite{rew1990unidata} & \textbf{(P)HDF5} \cite{folk1999hdf5} & \textbf{ADIOS} \cite{lofstead2008flexible}  \\
        \midrule
        \textbf{open}  & \texttt{nc(mpi)\_open}            & \texttt{H5Fopen}  & \texttt{adios\_open (r)}           \\ 
        \textbf{create}  & \texttt{nc(mpi)\_create}            & \texttt{H5Fcreate}  & \texttt{adios\_open (w)}           \\ 
        \textbf{read}  & \texttt{nc(mpi)\_vara\_get\_type} & \texttt{H5Dread}  & \texttt{adios\_schedule\_read} \\ 
        \textbf{close} & \texttt{nc(mpi)\_close}           & \texttt{H5Fclose} & \texttt{adios\_close}          \\
        \bottomrule
        \vspace{0.01em}
    \end{tabular}
    \caption{Mapping data access operations to I/O libraries \vspace{-2em}}
    \label{table:opmap}

    \normalsize
\end{table} 
%

%

\subsubsection{\abbr{} APIs}
\abbr{} provides an additional API providing more information and control
about the virtualized environment. 
%
%
These functions do not perform I/O: they are issued before
the I/O calls to coordinate with the DV before accessing the files.

\textbf{Initialize/Finalize: }
An analysis tool can start an analysis on a given simulation context
by calling the \texttt{SIMFS\_Init} function. Multiple contexts can
be open by the same application.
%
\begin{lstlisting}
int SIMFS_Init(char * sim_context, SIMFS_Context * context);
int SIMFS_Finalize(SIMFS_Context * context);
\end{lstlisting}

\textbf{Requesting Data: }
Before accessing a set of files with standard I/O libraries, the analysis
acquires such files with the \texttt{SIMFS\_Acquire} function. This function
blocks until the DV notifies that the requested files are available. A
non-blocking version of the call is available that does not wait for the
requests files to become available: the application must then explicitly test
or wait for data availability.
\begin{lstlisting}
int SIMFS_Acquire(SIMFS_Context context, char * filenames[], 
    int count, SIMFS_Status * status);
int SIMFS_Acquire_nb(SIMFS_Context context, 
    char * filenames[], int count, SIMFS_Status * status, 
    SIMFS_Req * req);
int SIMFS_Release(SIMFS_Context context, char * filename);
\end{lstlisting}
The acquire functions return a \texttt{SIMFS\_Status} object containing
information such as the error state (e.g., restart failed) and the
estimated waiting time for the requested files to become available.  The
analysis can use this information for debugging, profiling, and for saving
compute hours/energy (e.g., by checkpointing itself and requesting to be
resumed after the estimated waiting time).
Once the analysis of a file finishes, the application releases it
with a \texttt{SIMFS\_Release} call.

\textbf{Waiting for Data: }
The application can wait or test for the completion of non-blocking acquire
calls with the \texttt{SIMFS\_Wait} and \texttt{SIMFS\_Test} functions,
respectively. These functions return a \texttt{SIMFS\_Status} object to inform
the application about the status of the re-simulation.  Since an acquire
request can target multiple files with different states (i.e., on disk or
missing), we provide the \texttt{SIMFS\_Waitsome} and \texttt{SIMFS\_Testsome}
calls that allow to receive availability information for a subset of files
requested in the acquire call.

\begin{lstlisting}
int SIMFS_Wait(SIMFS_Req * req, SIMFS_Status * status);
int SIMFS_Test(SIMFS_Req * req, int * flag, 
    SIMFS_Status * status);
int SIMFS_Waitsome(SIMFS_Req * req, int * readycount, 
    int readyidx[], SIMFS_Status * status);
int SIMFS_Testsome(SIMFS_Req * req, int * readycount, 
    int readyidx[], SIMFS_Status * status);
\end{lstlisting}

\textbf{Comparing Data: } 
If bitwise reproducibility is not guaranteed, the analysis can check if a given
file matches the one produced by the initial simulation with the
\texttt{SIMFS\_Bitrep} call. The check is made by comparing the checksums of
the current file and the original one. The way the checksum is computed is
simulator-specific and specified as a function of simulator driver. The
simulation context keeps a map from filenames to checksums that can be updated
through a command line utility at the time when the first simulation is run.
\begin{lstlisting}
int SIMFS_Bitrep(SIMFS_Context context, char * filename, 
    int * flag);
\end{lstlisting}


\subsection{Caching Simulation Data} \label{sec:caching}

Simulation data virtualization is sufficient to fully solve data storage
limitations because it allows to freely adjust the space-time-tradeoff by
re-creating data on demand.
Yet, re-simulating every file may be too slow and with limited disk-space, it
is unclear which files should remain on disk and which should be re-created on
demand.

Traditional caching theory classifies cache misses using the 3Cs
model~\cite{Hill:2014:CCA:2692916.2558890} as compulsory, capacity, and
conflict misses. In our model, we first run a whole simulation to create
restart files and these initial compulsory misses cannot be avoided. 
Conflict misses are caused by low-latency caching schemes
that map blocks to sets to optimize the performance. Since our system
is operating on a milliseconds time-frame, we employ fully
associativity, avoiding conflict misses. However, if the data
does not fit in cache, we may need to evict files from the cache due
to the limited storage, causing capacity misses. 

Caching simulation data is different from caching memory accesses in system
caches: 
here, a cache miss leads to the re-simulation of a number of output steps which
depends on the restart interval and the missing output step.
Also, the replacement schemes need to take into account that may not be
possible to evict some output steps if they are currently referenced by one or
more analyses.
%
We now discuss a set of known replacement schemes that we extend to fulfill the
requirements for simulation data virtualization. 

\textbf{Locality-Based (LRU/LIRS/ARC): }
Least-Recently-Used (LRU) is one of the most common and simplest replacement
schemes. The idea is to keep track of the recency of each cache entry (i.e.,
how many accesses have been issued from the last access to it) and select the
least recently used one as victim. 
More advanced locality-based schemes have been proposed with the aim of 
improving over LRU. The key change is in how locality is defined (LRU 
defines it as recency). 
Low Inter-reference Recency Set (LIRS)~\cite{jiang2002lirs} leverages both 
recency and reuse distance (i.e., number of accesses between two consecutive
accesses targeting the same entry) for selecting entries to replace.
Instead, Adaptive Replacement Cache (ARC)~\cite{megiddo2003arc} distinguishes 
entries that are frequently used from the ones that have been recently 
accessed: they are kept in two different sets which size is adjusted at
runtime in order to adapt to the observed access pattern. 

\textbf{Cost-Aware (BCL/DCL): }
The \emph{Basic Cost-Sensitive LRU} (BCL) and \emph{Dynamic Cost-Sensitive LRU}
(DCL) replacement schemes have been proposed by Jeong et
al.~\cite{jeong2006cache}.
The main idea is that they do not evict the LRU if there is a more recent entry
with a lower miss cost: the victim is selected as the first entry in the
recency-ordered list with a cost lower than the one of the LRU. LRU is used as
fallback if no evictable entry can be find in this search.
If the LRU is not evicted, its cost gets reduced to avoid the case in which a
costly, sporadically-accessed entry leads to the eviction of too many cheaper,
highly-reused entries.
In this context, the miss cost of an entry (i.e., output step) is the distance,
in number of output steps, from its closest previous restart step.
BCL and DCL differ by the time at which the LRU depreciation takes place: BCL
depreciates it as soon as the LRU is not evicted, while DCL does that only if
an evicted non-LRU entry gets accessed before the LRU. 
Jeong et al. propose also the \emph{Adaptive Cost-Sensitive LRU} (ACL) but we
choose to not consider this algorithm since it is not designed for fully
associative caches. 
\begin{figure}[h]
    \vspace{-1em}
    \centering{
        \includegraphics[trim={10 6 10 0}, clip, width=1\columnwidth]{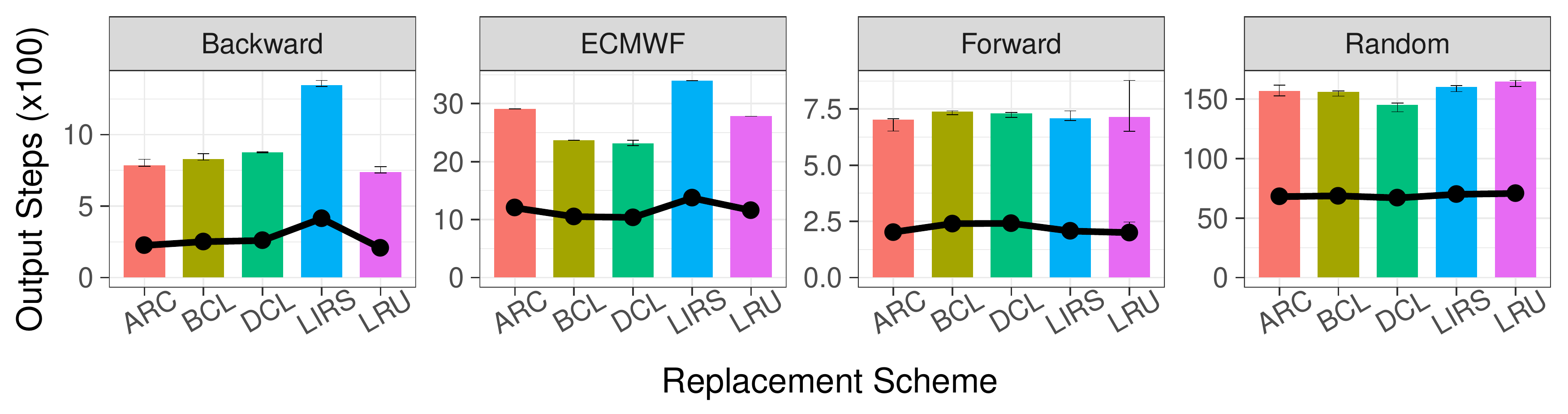}}
    \vspace{-2em}
    \caption{Cache replacement schemes comparison for different access 
             patterns.}
    \label{fig:cachecomp} 
    \vspace{-0.4em}
\end{figure} 

\textbf{Caching Schemes Evaluation: }
To evaluate the discussed caching schemes we virtualize a 4-days simulation
producing an output step every $5$ minutes and a restart
file every $4$ hours.  The \abbr{}'s cache is set to $25\%$ of the data volume.

The simulation data is accessed by a synthetic analysis tool that replicates a
given access trace.  We generate different traces for different analysis
patterns: \emph{forward}, where a set of output steps are accessed on a
forward-in-time trajectory; \emph{backward}, where the output steps are accessed
on backward-in-time trajectory; \emph{random}, where the accessed output steps
are randomly selected.
For each access pattern, we generate $50$ traces starting their analysis
at a random point of the simulation timeline and accessing a different
numbers of output steps (randomly selected between $100$ and $400$). 
We then concatenate all the single traces in a single one to be replicated
by our synthetic analysis tool.
In addition, we extract traces from the ECMWF
archive~\cite{Grawinkel:2015:AES:2750482.2750484} that provides a complete
trace of all successful accesses to the ECFS archival system from January 2012
to May 2014.  The resulting trace accesses $874$ different files for a total of
$659,989$ times. 

Figure~\ref{fig:cachecomp} shows the re-simulation statistics: the bars
represent the number of simulated output steps for the different replacement
schemes (x-axis) and different access patterns (tiles). We also report the
number of times a new simulation has been restarted to satisfy the analysis
(black points). 
We repeat each experiment $100$ times, generating new traces each time, and
report the median and the 95\% CI of the measured counts.
Except for LIRS, we notice no important differences among the caching schemes
for scan-like access patterns (i.e., forward and backward).  LIRS performs
worse in the backward case because it prioritizes the eviction of files that
are most likely to be accessed with this trajectory.
The cost-based schemes, in particular DCL, minimize the number or
restarts/produced output steps in the ECMWF and random cases. Since multiple
analysis tools accessing data with different access patterns can be interfaced
to \abbr{} at the same time, we expect that the random and ECMWF traces to be
the most similar to real-word scenarios. Hence, in the following, we fix the
caching replacement scheme to DCL.

\subsection{Virtualizing Simulation Pipelines}

Many scientific simulation are organized in stages: e.g., the initial boundary
conditions are copied from long-term storage to start a coarse-grain simulation
that outputs data that is then used as input of a finer-grain simulation.  
If we virtualize the fine-grain simulation output we may need to re-simulate parts
of it, needing the output of the coarser-grain one. However, storing all
the output of the coarse-grain simulation to re-simulate any portion of the
fine-grain one may be prohibitive, leading us to our initial problem (i.e., we
cannot keep all the data on-disk).
\begin{figure}[h]
    \vspace{-1em}
    \centering{
        \includegraphics[trim=20 14 20 10, clip, width=0.9\columnwidth]{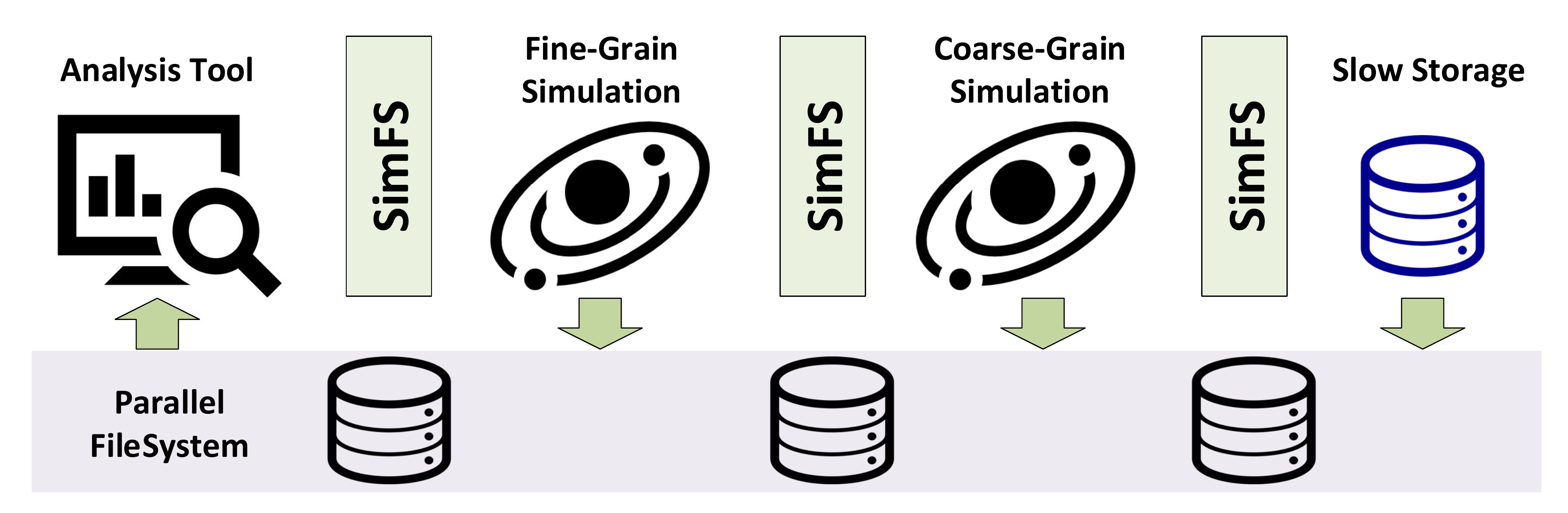}}
    \vspace{-0.5em}
    \caption{Using \abbr{} to virtualize simulation pipelines}
    \label{fig:pipeline}
    \vspace{-0.5em}
\end{figure}
We have two options to address this problem: 1) the simulation job of the
fine-grain simulation makes sure that all the needed input is on disk (e.g., by
starting the coarser-grain simulation first or by copying it from long-term
storage); 2) we virtualize the output of all the stages, as shown in
Figure~\ref{fig:pipeline}. In the second case we define a simulation context
for each stage: if the fine-grain simulation accesses a part of its input that
is missing, then a coarse-grain re-simulation will be started.  Similarly, if
the coarser-grain simulation accesses missing parts of its input, then a new
simulation job will be created: in this case, this job will not start a
simulation but just issue the copy of the data from the long-term storage area.

\section{Optimizing Simulation Data Accesses} 

Many analysis tools access the data with simple traversal schemes such as
forward or backward in time trajectories. These access patterns can be
optimized using prefetching strategies that can hide the simulation startup
latency as well as improve the overall production bandwidth.
%
%
These prefetching strategies can be used to adjust another resource-performance
tradeoff. For example, in the common case where the simulation produces data
slower than the analysis tool can consume it, we can use more resources to run
many simulations in parallel and match the analysis application ingestion
bandwidth. 
%
%

%
%

\subsection{Performance Model}\label{sec:performancemodel}


We start by defining a performance model for the simulations and the analysis
applications that is then used in the proposed optimizations. The idea is to
have a general performance model that allows us to not make particular
assumptions on the simulator and analysis.

Restarting a simulation may incur in non-functional delays such as waiting for
resources (e.g., VM deploying or queuing time in a batch system), reading
the restart file, and initializing the simulation model. 
We define $\alpha_{\text{sim}}(p)$ as the restart latency of a simulation
running with parallelism level $p$.
%
Once started, the simulation writes the output step on disk with a certain
frequency: we model the simulation inter-production time as
$\tau_{\text{sim}}(p)$, that represents the time between the production of two
consecutive output steps.  
In the following, we omit $p$ for both $\alpha_{\text{sim}}(p)$ and
$\tau_{\text{sim}}(p)$ if not required by the context.
According to this model, the time needed to simulate $n$ output steps using a
parallelism level $p$ is: $ T_{\mathrm{sim}}(n, p) = \alpha_{\mathrm{sim}}(p) +
n \cdot \tau_{\mathrm{sim}}(p) $. Hence, the time to produce an output step
$d_i$ is the simulation time from $R(d_i)$ to $d_i$ itself:
\mbox{$T_{\mathrm{sim}}(i - R(d_i), p)$}.

We model the analysis application performance as $\tau_{\text{cli}}^k$, that is
the time between two consecutive k-strided accesses. 


\subsection{Prefetching Simulation Data}\label{sec:prefetching}

%

We associate each analysis application that is interfaced to \abbr{} with a
prefetch agent. The prefetch agent monitors the application access pattern,
measures $\tau_{\text{cli}}^k$, and can prefetch new re-simulations.  Forward
and backward access patterns are detected after two $k$-stride consecutive
accesses.  Once a pattern is detected, the agent starts prefetching
re-simulations according with the monitored parameters. A prefetch agent resets
itself whenever the analysis tool changes its analysis direction and/or stride,
or terminates.

%

\subsubsection{Prefetching forward-in-time accesses} 

We start with the simplest and most common pattern: forward-in-time. This
pattern is directly supported by in-situ, where the analysis tool runs in
tandem with the simulation. While a single simulation with in-situ analysis is
always faster than re-simulation, \abbr{} has many benefits if the data needs
to be analyzed at varying times (e.g., by different analysis). In fact, we can
improve this scenario at two fronts: (1) we can use all the storage available
to cache output steps for future analyses and (2) we can reduce the analysis 
completion time using prefetching.

\paragraph{\textbf{Masking Restart Latency}}\label{sec:masklat}

A forward-in-time analysis reads the files in the same time trajectory they are
produced by the simulation. If no prefetching strategies are adopted, \abbr{}
starts a new simulation only when a miss occurs, making the analysis application
wait the full restart latency at every miss.
\begin{figure}[h]
    \vspace{-1.5em}
    \centering{ 
        \includegraphics[trim={0 425 500 10}, clip, width=1\linewidth ]{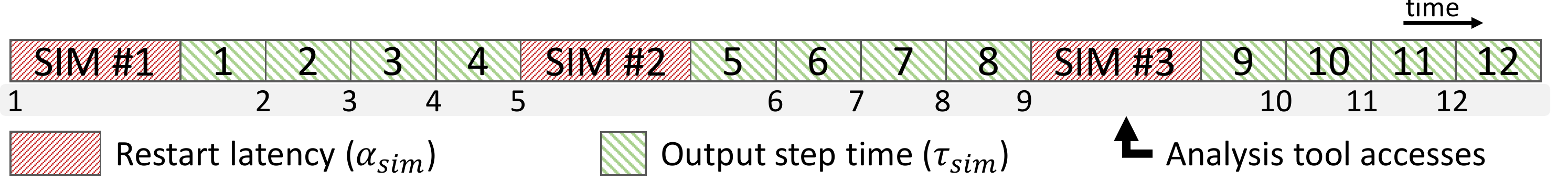}}
    \vspace{-1.8em}
    \caption{Forward analysis without prefetching.}
    \label{fig:no_prefetching}
    \vspace{-1.2em} 
\end{figure} 

\begin{figure}[h]
    \centering{
    \includegraphics[trim={0 420 500 0}, clip, width=1\linewidth ]{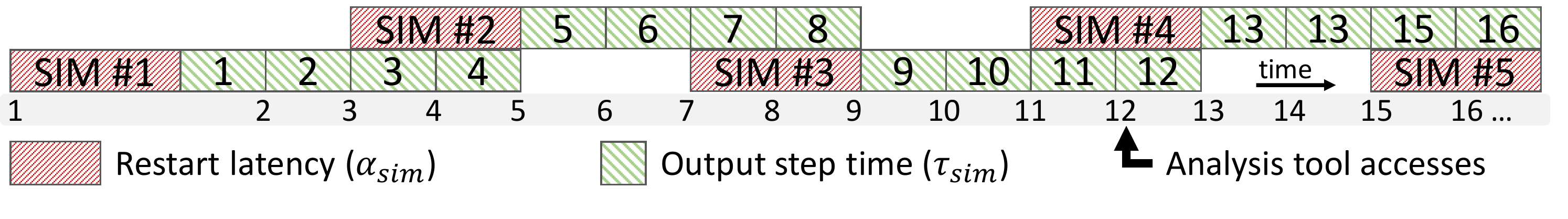}}
    \vspace{-1.8em}
    \caption{Hiding simulation restart latency.}
    \label{fig:pref_mask} 
    \vspace{-1.9em}
\end{figure}

Figure~\ref{fig:no_prefetching} shows an example of an analysis making a
sequence of ($k=1$)-strided accesses with all of them resulting in a miss.  The
simulation has a restart interval of $\Delta r = 4$ timesteps, the restart
latency is $\alpha_{\text{sim}} = 2$ time units, and produces one output step
every time unit ($\tau_{\text{sim}} = 1$ time unit).  The analysis consumes the
output steps twice as fast as they are produced ($\tau_{\text{cli}}^k = 1/2$ time
units). The accessed output steps are reported into the gray bar at the bottom. 
The example shows how the accesses performed by the analysis are delayed of the
the restart latency every time a miss occurs.
We want to mask the restart latency by overlapping it with the analysis, as
shown in Figure~\ref{fig:pref_mask}. This leads to two questions: \emph{How
long does the re-simulation need to be?} and \emph{When to trigger a new
re-simulation?}

The re-simulation length $n$ is the number of output steps that one
re-simulation produces. The number of $k$-stride accesses that can be served by
one re-simulation is $\lfloor \frac{n}{k} \rfloor$. 
The analysis processing time per output step is \mbox{$\max(k \cdot
\tau_{\text{sim}}, \tau_{\text{cli}}^k)$}: it can be limited by either the
simulation's or its own speed. We want to find an $n$ such that the time spent
in analyzing $\lfloor \frac{n}{k} \rfloor$ output steps covers the restart
latency of the next re-simulation, reserving the first two accesses to confirm
the prefetching validity (i.e., same direction and stride). This $n$ can be
found by satisfying the following inequality: \vspace{-1em}

\[ \left(\Bigl\lfloor \frac{n}{k} \Bigr\rfloor - 2 \right) \cdot \max(k \cdot \tau_{\text{sim}}, \tau_{\text{cli}}^k) \ge \alpha_{\text{sim}} \] \vspace{-1em}

Hence, $n$ needs to be: 
$n \ge \Bigl\lceil\frac{\alpha_{\text{sim}}}{\max(k \cdot \tau_{\text{sim}}, \tau_{\text{cli}}^k)} + 2 \Bigr\rceil \cdot k$.
We always round $n$ up to the nearest restart interval multiple: \vspace{-1em}

\[ n = R\left( \Bigl\lceil\frac{\alpha_{\text{sim}}}{\max(k \cdot \tau_{\text{sim}}, \tau_{\text{cli}}^k)} + 2 \Bigr\rceil \cdot k + \frac{\Delta r}{\Delta d} \right) \]\vspace{-1em}

The abstraction we want to provide to the analysis tool is as if there is a
single simulation serving all the non-cached output steps it requests.  Hence,
we need to prefetch a new re-simulation just in time to mask its restart
latency.
Since the prefetch agents see the time as discretized by the (strided) analysis
accesses, we prefetch at the time of the last $k$-strided access that allows the
masking of the restart latency. This output step, named \emph{prefetching step},
is computed as: 
\mbox{$ d_i + n - \Bigl\lceil \frac{\alpha_{\text{sim}}}{\max(k \cdot \tau_{\text{sim}}, \tau_{\text{cli}}^k)} \Bigr\rceil \cdot k $}, where $d_i$ is the initial output step of the
currently running simulation.

%

\paragraph{\textbf{Matching Analysis Bandwidth}}

While the hiding of the restart latency avoids delaying the analysis at every
miss, the analysis can still be faster in consuming the output steps than the
simulation in producing them: \mbox{$\tau_{\text{sim}}>\frac{\tau_{\text{cli}}^k}{k}$}.  In
this case, we can improve the simulation production bandwidth by using two
strategies:
(1) \emph{Increase the simulation parallelism level}, or (2)
\emph{start multiple simulations in parallel}.
%

Strategy (1) is the first strategy that a prefetch agent employs if the
analysis is faster than the simulation.  When the application is accessing the
output steps produced by a simulation (i.e., due to a miss), the prefetch agent
monitors both $\tau_{\text{cli}}^k$ and $\tau_{\text{sim}}(p)$: whenever the
analysis is faster than the simulator, the prefetch agent increases $p$ for the
next re-simulation that will be started to recover the misses of this analysis. 
Whenever the prefetch agents determines that there are no performance benefits
in increasing $p$ or the \emph{max parallelism level} is reached, it switches
to strategy (2).
%

\begin{figure}[h]
    \vspace{-1em}
    \centering{
        \includegraphics[trim={0 328 500 0}, clip, width=1\linewidth ]{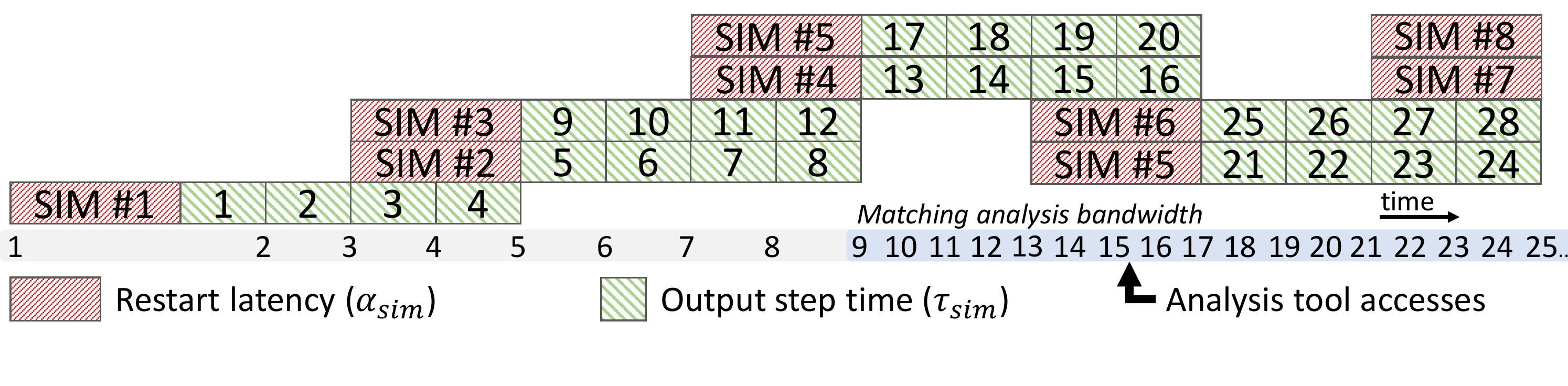}}
    \vspace{-2.6em}
    \caption{Hiding restart latency and matching forward analysis bandwidth.}
    \label{fig:pref_fw}
    \vspace{-0.7em}
\end{figure}

Strategy (2) runs multiple re-simulations in parallel to increase
the simulation output bandwidth. The ideal number of parallel re-simulations 
needed to match the analysis bandwidth is: $s_{\text{opt}}=\lceil
{k \cdot \tau_\text{sim}}/{\tau_\text{cli}^k} \rceil$. 
Figure~\ref{fig:pref_fw} shows how this strategy changes the example of
Sec.~\ref{sec:masklat}: the prefetch agent now starts $s_{\text{opt}} = 2$ new
re-simulations at each prefetching step and, after the first batch of
prefetched simulations (i.e., accessed output step $9$), the analysis can run
at its full bandwidth.
However, this strategy can lead \abbr{} to launch a large number of
re-simulations if the analysis is much faster than the simulation. Also, it is
not guaranteed that the prefetched output steps will be accessed by the
analysis, which can terminate or change its direction/stride at any time.  To
limit this issue, a simulation context can be configured to not prefetch
directly $s_{\text{opt}}$ simulations at time, but start with $s=1$ and double
it at each prefetching step until the analysis stays on the same
direction/stride and $s < \min(s_{\text{opt}}, s_{\text{max}})$, where $s_{\text{max}}$
is a simulation context parameter that limits the maximum number of simulations
that can be running at the same time.

\subsubsection{Prefetching backward-in-time accesses}\label{sec:back_pref}

Backward-in-time accesses are common in root-cause analysis. They are
conceptually similar to forward-in-time but require a different prefetching
scheme because the simulation itself is always forward-in-time.  Because of
this, the analysis cannot operate in tandem with the simulation (like in-situ):
if $d_i$ is missing, the analysis has to wait until the
re-simulation produces the output steps from $R(d_i)$ to $d_i$, like in
forward-in-time trajectories. However, since the analysis goes backward, now it
can find other output steps produced in that interval already in cache.
The output steps produced after $d_i$ (i.e., from $d_{i+1}$ to $R(d_i
+ \Delta r)$) are not useful to the analysis. 
Hence, prefetching for backward-in-time analysis requires to mask not only the
restart latency but also (part of) the re-simulation itself.  

Let us consider the case where the analysis is slower than the simulation:
$\frac{\tau_\mathrm{cli}^k}{k} > \tau_\mathrm{sim}$. 
To hide the re-simulation time, we have to simulate enough output steps such
that the time the analysis needs to consume them is higher than the cost of the
next simulation: \mbox{$ \frac{n}{k} \cdot \tau_\mathrm{cli}^k  \ge
\alpha_\mathrm{sim} + n \cdot \tau_\mathrm{sim}$}.  Hence, the minimum number
of output steps to be simulated is: \mbox{$n = \frac{k \cdot
\alpha_\mathrm{sim}}{\tau_\mathrm{cli}^k - k \cdot \tau_\mathrm{sim}}$}, 
rounded up to the next restart step.

If the analysis is faster than the simulation, then we have again two
strategies: increase the simulation parallelism or run multiple simulations in
parallel.
The ideal number of parallel re-simulation we need to match a backward-in-time
analysis bandwidth is different from the forward-in-time case. Here, once a
missing output step is produced, the analysis can find in cache all the next
output steps on its trajectory up to the restart step used for the last
re-simulation. Hence, we want to produce a number of output steps such that
the time the analysis takes to process them (at its full speed) is 
greater than the time to prefetch a new set of output steps:

\vspace{-0.7em}
\[ s \cdot \frac{n}{k} \cdot \tau_\mathrm{cli}^k \ge \alpha_\mathrm{sim} + n \cdot \tau_\mathrm{sim} \] 
\vspace{-1.3em}

\noindent hence, the minimum number of parallel simulations is: 

\vspace{-0.7em}
\[ s = \frac{k \cdot \alpha_\mathrm{sim}}{n \cdot \tau_\mathrm{cli}^k} + \frac{ k \cdot \tau_\mathrm{sim}}{\tau_\mathrm{cli}^k} \]
\vspace{-0.7em}



\begin{figure}[t]
    \centering{
        \includegraphics[trim={0 385 500 0}, clip, width=1\linewidth ]{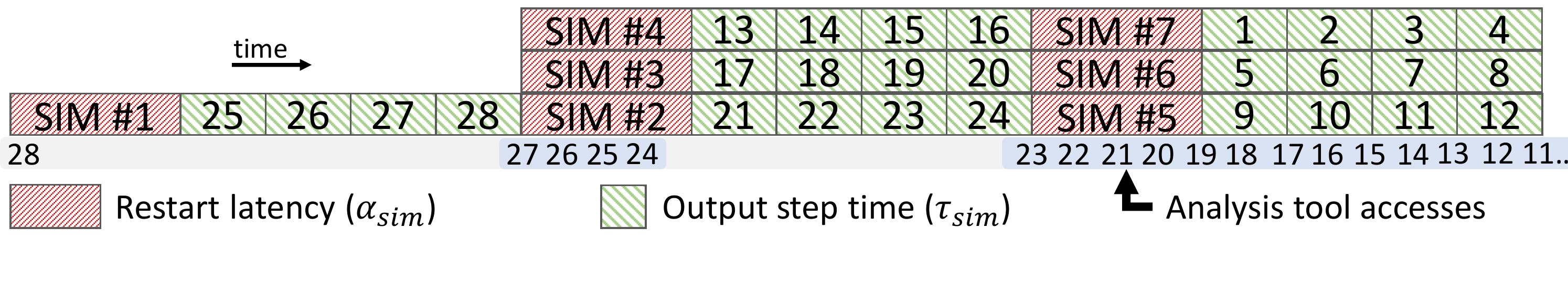}}
    \vspace{-2.6em}
    \caption{Hiding restart latency and matching backward analysis bandwidth.}
    \label{fig:pref_bw} 
    \vspace{-1.8em}
\end{figure}

%
This introduces a trade-off between $s$ and $n$: the higher the multiple
parallel simulations ($s$). the lower the number of output steps per simulation
($n$) that are needed to match the backward-in-time analysis bandwidth. 
However, reducing $n$ by using more computing resources in parallel allows us
to reduce the time needed to reach the full bandwidth. 
Figure~\ref{fig:pref_bw} shows a backward analysis with $\alpha_{\text{sim}} =
2$, $\tau_{\text{sim}} = 1$, $\tau_{\text{cli}}^k = 1/2$, $k=1$, and $n=4$. In
this case, the minimum number of parallel resimulation needed to match the
analysis bandwidth is $s=3$. The example shows how a new batch of re-simulations
(\texttt{SIM \#5,6,7}) can be overlapped to the analysis of the output steps
produced by the previous one (\texttt{SIM \#2,3,4}).

\enlargethispage*{\baselineskip}

\subsection{Prefetching Effectiveness}

The discussed prefetching strategies aim to hide the restart overhead and
increase the overall simulation bandwidth. 
However, to avoid a too aggressive prefetching that would lead to cache
pollution, \abbr{} tries to kill simulations prefetched by analyses that
terminated or changed analysis direction (e.g., from forward to
backward or jumped to a different timespan). A simulation can be killed only if
there are no other analyses waiting for the files that are going to be produced
by it. Additionally, \abbr{} tries to detect cache pollution by monitoring
the accesses to the prefetched output steps: if an analysis accesses an output
step that has been prefetched by the prefetch agent associated with it and
finds it missing, this means that this file has been produced and evicted
before being accessed: this is considered a cache pollution signal and leads
to the reset of all the active prefetch agents.

\subsubsection{Prefetching with high restart latencies} \label{sec:high_lat}

Before producing their effects (i.e., masking the restart latency and matching
the analysis bandwidth) the prefetching strategies need a warm-up period of
time. The warm-up length depends, among the others, on the restart latency,
which includes the system overheads for restarting re-simulations (e.g.,
queuing time in a batch system).  The overheads can vary according with the
system where \abbr{} is deployed (e.g., cloud or HPC systems). We now quantify
this warm-up time and discuss its effects on the prefetching effectiveness. 
For simplicity, we assume an empty \abbr{} cache, and a single running analysis
accessing $m$ output steps with stride $k=1$. 
%


\begin{center}
    \centering{
        \includegraphics[trim={0 358 500 5}, clip, width=1\linewidth ]{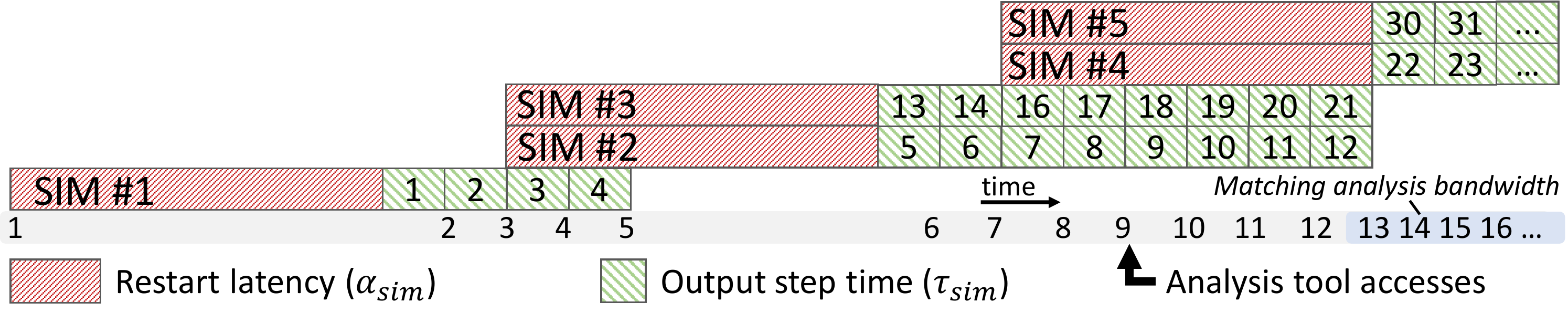}}
    \vspace{-1.8em}
    \captionof{figure}{Prefetching with high restart latencies.}
    \label{fig:pref_hlat}
    \vspace{-0.2em}
\end{center}

\enlargethispage*{\baselineskip}

\paragraph{Forward-in-time prefetching}
Let us define $\alpha_{\text{sim}}^i$ as the restart latency experienced by the $i$-th
re-simulation. We initially assume a constant
restart latency: \mbox{$\alpha = \alpha_{\text{sim}}^i = \alpha_{\text{sim}}^j \, \forall i,j$}.
Figure~{\ref{fig:pref_hlat}} shows an example of how the restart latency
can impact the prefetching effectiveness.
When the analysis accesses the first missing output step, a full restart
latency is paid. After this, the re-simulation starts producing output steps
every $\tau_{\text{sim}}$ time units. Recall that, at time of the first
miss, \abbr{} has no information about the analysis access pattern, hence a
single restart interval is simulated (i.e., $\frac{\Delta r}{\Delta d}$ output
steps). 
Once the next two output step are requested by the analysis, \abbr{} can
determine the analysis direction and prefetch a new set of $s$ re-simulations.
The maximum part of restart latency of these re-simulations that can be
masked is \mbox{$\frac{\Delta r}{\Delta d} \cdot \tau_{\text{sim}}$} time units
(assuming the simulation is slower than the analysis: \mbox{$\tau_{\text{sim}} >
\tau_{\text{cli}}^k$}). Each of these re-simulations produce a number $n$ of output
steps that will be enough to cover the next restart latencies (see
Sec.~{\ref{sec:masklat}}).
The effects of the prefetching impact the analysis performance only after the
second simulation finishes (i.e., \texttt{SIM \#2}). At this time, the analysis
will need the output steps produced by the third re-simulation, that will now
be in cache.

Summing up, the warm-up time, $T_{pre}^{fw}$, can be defined as: $T_{pre}^{fw} = \alpha_{\text{sim}}
+ \max(2 \cdot \tau_{\text{sim}} + \alpha_{\text{sim}}, \frac{\Delta r}{\Delta d} \cdot \tau_{\text{sim}})
+ n \cdot \tau_{\text{sim}} $. Assuming $\tau_{\text{sim}} \ll \alpha_{\text{sim}}$ and $\frac{\Delta r}{\Delta d} \cdot
\tau_{\text{sim}} < \alpha_{\text{sim}}$, this can be approximated with: $T_{pre}^{fw} \approx 2 \cdot
\alpha_{\text{sim}} + n \cdot \tau_{\text{sim}}$. After the prefetching warm-up, \abbr{} will be
able to always mask the restart latencies, producing output steps every
$\frac{\tau_{\text{sim}}}{s}$ time units on average. Hence, the analysis time can be
defined as (assuming $m>n$):

\vspace{-0.8em}
\[ T_{cli}^{fw} \approx T_{pre}^{fw} + (m-n) \cdot \frac{\tau_{\text{sim}}}{s} =  2 \cdot \alpha + n \cdot \tau_{\text{sim}} + (m-n) \cdot \frac{\tau_{\text{sim}}}{s}.\]
\vspace{-1.3em}

\noindent This shows an Amdahl's law effect on the prefetching strategies
scalability: the higher the restart latency, the longer the prefetching warm-up
(where no speedup can be seen).  This can be compensated by longer analysis
(i.e., large $m$), that can make the sequential part negligible.

\paragraph{Backward-in-time prefetching}
Backward-in-time analysis experience higher prefetching warm-up times. 
As for the forward-in-time case, a full restart latency is paid when the
first missing access is made by the analysis (namely $d_i$). Recall that the restarted
simulation can only go forward in time: The second access (which
will determine the analysis direction) can be made only after the first
\mbox{$D_i = d_i - R(d_i)$} output steps are produced and the first
missing output step is analyzed (taking $\tau_{\text{cli}}^k$ time units). 
After the analysis direction is determined, \abbr{} can start prefetching 
re-simulations as described in Sec.~\ref{sec:back_pref}. The effects of
the prefetching on the analysis time will be visible only after the first
batch of prefetched simulations will be complete. We can define the warm-up
time for backward-in-time prefetching as: 
$T_{pre}^{bw} = \alpha_{\text{sim}} + D_i \cdot \tau_{\text{sim}} + \tau_{\text{cli}}^k + 
\max(\tau_{\text{cli}}^k \cdot (D_i - 1), \alpha_{\text{sim}} + n \cdot \tau_{\text{sim}})$.
Assuming an analysis faster than the simulation (i.e., $\tau_{\text{sim}} > \tau_{\text{cli}}^k$)
and being $n \ge \frac{\Delta r}{\Delta d} \ge D_i$, we can approximate it
as $T_{pre}^{bw} \approx 2 \cdot \alpha_{\text{sim}} + D_i \cdot \tau_{\text{sim}} + n \cdot \tau_{\text{sim}}$.
Differently from the forward-in-time case, here the prefetching warm-up accounts for
the $D_i$ value, which depends on where the analysis starts (i.e., $d_i$) and
the restart interval.

\paragraph{Non-constant restart latencies} If the restart latencies are not
constant (e.g., high variability of the jobs queueing times), \abbr{} may not
be able to always mask the restart latencies.  
To account for this case, \abbr{} keeps track of the restart latencies using an
exponential moving average, so to consider only the most recent observation (the
smoothing factor is a parameter defined in the simulation context). Whenever
the restart latency is underestimated, the analysis is delayed by this
estimation error. If we define $A$ as the sequence of re-simulation serving the
requests of an analysis and $\bar{\alpha}_{sim}^i$ as the restart latency
estimation for the $i$-th re-simulation, we can quantify this additional delay
as: $ \sum_{i \in A} \max(0, \alpha_{\text{sim}}^i - \bar{\alpha}_{sim}^i) $.

\section{Cost Analysis}\label{sec:costmodel}

We now introduce cost models for the different simulation data analysis 
solutions: on-disk, in-situ, and \abbr{}. We use these models for studying the
cost-effectiveness of the different solutions. 
We assume that the data needs to be made available for analyses for a
fixed period of time, that we call \emph{simulation data availability
period} $\Delta t$.
During this period of time, the data is either stored on disk in the
on-disk method; simulations are started for each analysis in in-situ;
or data is virtualized via \abbr{}. 
We assume that the simulation cost does not include the restart latency
$\alpha_{\text{sim}}$, that is the non-billed waiting time before the
simulation job actually starts running (e.g., VM deploying time or job queueing
time in a batch system). 

We identify two main costs: the storage $c_s$ and computation $c_c$ costs. The
first accounts for the monthly storage of one GiB of data ($\$/GiB/month$); the
second for one hour of computation on a single compute node ($\$/node/hour$).
The output and restart steps sizes (GiB) are assumed to be constant: they are
represented with $s_o$ and $s_r$, respectively. The number of output
steps and restart steps produced by a simulation of $n$ timesteps are 
\mbox{$n_o = \left\lfloor \frac{n}{\Delta d} \right\rfloor$} and 
\mbox{$n_r = \left\lfloor \frac{n}{\Delta r} \right\rfloor$}, respectively.

We now define the costs of simulating and storing a number of output steps,
that are the building blocks of the cost models discussed below.
Simulating $O$ output steps using $P$ compute nodes has cost 
$ C_{\text{sim}}(O, P) = {O \cdot \tau_{\mathrm{sim}}(P) \cdot P \cdot c_c}$:
%
This is the time to produce a single output steps using $P$ nodes times the
number of output steps to produce, times the hourly compute cost.
Storing $F$ files of size $s$ for $\Delta t$ months has cost
$ C_{\text{store}}(F, m, \Delta t) =  F \cdot m \cdot \Delta t \cdot c_s $:
this is cost of storing $1GiB$ of data for $\Delta t$ months times the 
file size (in $GiB$).
Table~\ref{table:cost_sym} summarizes the symbols used by this cost model.

\textbf{On-disk: }
This solution executes the full simulation and stores the output
for the entire data availability period.  This cost is independent
of the analyses that are performed on the simulation data.  It can be
expressed as the cost of the initial simulation plus the storage of $n_o$
output steps for $\Delta t$ months: 

\vspace{-1.5em}
\[ C_{\text{on-disk}}(\Delta t) = 
        C_{\text{sim}}(n_o, N) + 
        C_{\text{store}}(n_o, s_o, \Delta t) \] 
\vspace{-1.5em}

\textbf{\abbr{}}
%
Let us define the sequence of output steps that are accessed by all the
analyses performed during $\Delta t$ as $\gamma_{\Delta t}$ and let
$\gamma_{\Delta t}(j)$ be the subsequence of accesses made by an analysis $j$.
The number of output steps resimulated by \abbr{} when the sequence
$\gamma_{\Delta t}$ is observed is $V(\gamma_{\Delta t})$. This number depends
on the following factors: the restart interval $\Delta r$; the number and the
type of analyses performed; the cache size $M$ and its replacement policy; the
employed prefetching strategies.
We express the cost of enabling analysis of  simulation output over
$\Delta t$ months with \abbr{} as: 

\vspace{-1.5em}
\begin{multline*}
C_{\text{SimFS}}(\Delta t) = 
        C_{\text{sim}}(n_o, P) +
        C_{\text{store}}(n_r, s_r, \Delta t) + \\
        C_{\text{store}}(M, s_o, \Delta t) + 
        C_{\text{sim}}(V(\gamma_{\Delta t}), P)
\end{multline*}
\vspace{-1.5em}

\noindent This cost accounts for: the initial simulation (that produces the
restart steps); the storing the restart steps and the cached output
steps; and the re-simulation of the missing output steps.


\textbf{In-situ}
In-situ always couples a simulation with a running analysis. 
Let us assume an analysis $j$ accessing $|\gamma_{\Delta t}(j)|$ output steps and starting 
from the output step with index $i_j$ in a forward in time direction. With in-situ, 
this analysis requires a simulation from output step $d_0$ until $d_{i_j+|\gamma_{\Delta t}(j)|}$. 
Note that the output steps $d_0 \dots d_{i_j-1}$ are not useful to the analysis.
Enabling in-situ analysis for $\Delta t$ months has cost: 

\vspace{-1em}
\[ C_{\text{in-situ}}(\Delta t) = \sum_{j=1}^{z} C_{\text{sim}}(i_j + |\gamma_{\Delta t}(j)|, P) \]

\vspace{-0.5em}
\noindent where $z$ is the number of analyses performed during $\Delta t$. 

\begin{table}
\footnotesize
\centering
\begin{tabular}{>{\centering\arraybackslash}p{4em}|p{23em}}
\textbf{Symbol} & \textbf{Definition} \\
\toprule
$\Delta t$ & Simulation data availability period \\
$c_c$ & Compute cost ($\$/node/hour$) \\
$c_s$ & Storage cost ($\$/GiB/month$) \\
$n$   & Number of timesteps \\
$n_o$ & Number of output steps \\
$n_r$ & Number of restart steps \\
$s_o$ & Output step size (GiB) \\
$s_r$ & Restart step size (GiB) \\
$P$ & Number of compute nodes used to run re-simulations \\
\bottomrule
\end{tabular}
\vspace{-0.5em}
\caption{List of symbols introduced by the cost models}
\label{table:cost_sym}
\vspace{-2.7em}
\end{table}

\subsection{Cost-Effectiveness}
\label{sec:costeff}

We now use the cost models developed in Sec.~\ref{sec:costmodel} to compare
the costs of the standard analysis solution against \abbr{}.

We calibrate the cost models on the Microsoft Azure cloud platform because the
offered node types (NVIDIA Tesla P100 GPUs~\cite{azure}) are close to our
experimental settings: the \textbf{compute cost} is \mbox{$c_c =
2.07\$/node/hour$}. This is the hourly cost of a NCv2 virtual
machine~\cite{azure-vm-pricing}; the \textbf{storage cost} is \mbox{$c_s =
0.06\$/GiB/month$}, which is the monthly cost of storing $1GiB$ of data in an
Azure File share~\cite{azure-storage-pricing}.
While cheaper and slower cloud storage solutions are available (e.g., Azure
Blob Storage, Amazon Glacier), we choose to calibrate the model on a solution
providing file abstraction, such that it can be directly targeted by I/O
libraries (e.g., HDF5). 
%
%

The performance model is calibrated on a COSMO simulation executing on Piz
Daint, a Cray XC50 machine running at CSCS. COSMO is a climate model for
long-term simulations (see Sec.~\ref{sec:evaluation}). The simulation advances
with $20s$ timesteps and outputs one output step every $\Delta d=15$ timesteps.
The simulation is executed over $P=100$ compute nodes equipped with an NVIDIA
Tesla P100 GPUs, producing one output step every $20$ seconds:
$\tau_{\text{sim}}(100)=20s$.  The output step size is $s_o=6$ GiB, while the
restart step size is $s_r=36$ GiB.  The total data volume produced by this
configuration is $50$ TiB.

We use a number of synthetic analysis tools, accessing a sequence of output
steps with a forward-in-time trajectory.  Each of these sequences starts at a
randomly selected output step, so that analyses access different subsets of the
simulation output steps.  
These analysis can overlap in time and this overlap can affect the state
of the \abbr{} cache. We express the analysis overlap as the percentage of
accesses that an analysis performs without being interleaved with others'
execution.
If these sequences are known in advance and they can
be batched, then a single in-situ simulation is always the most cost-effective
solution. Instead, \abbr{} aims at a different scenario, where the analyses are
not known in advance and they need to be served in an on-demand fashion.
%
%





\begin{figure}[h]
    \centering{
    \vspace{-1.1em}
    \includegraphics[trim={0 0 0 0}, clip, width=1\columnwidth]{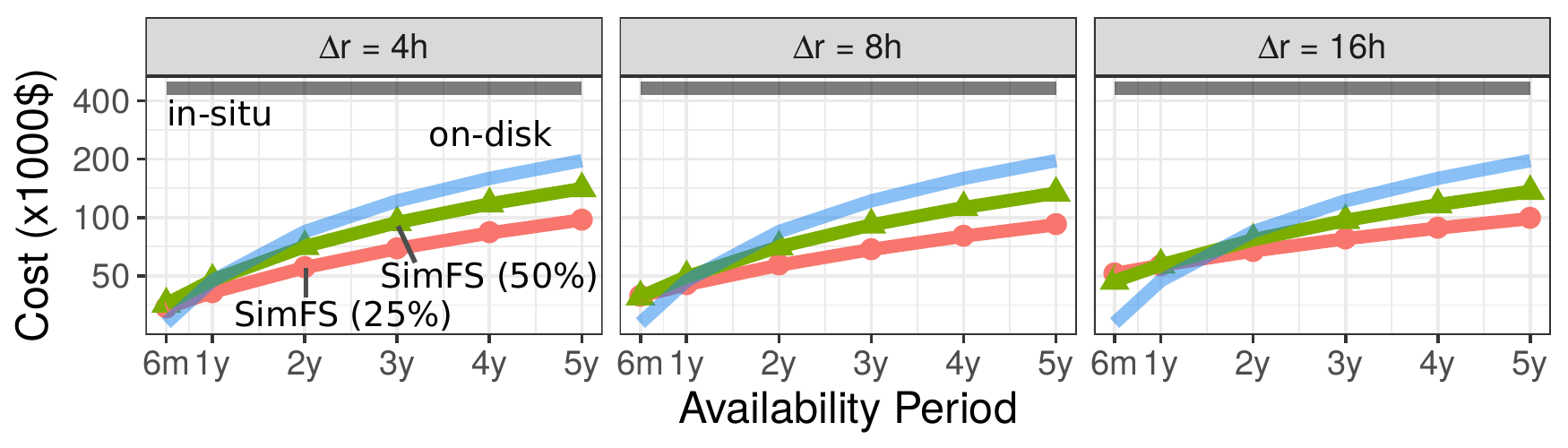}}
    \vspace{-1.9em}
    \caption{Data availability cost for different availability periods.}   
    \label{fig:costmodel_time} 
    \vspace{-1.0em}
\end{figure} 

\textbf{Simulation Data Availability Period}\label{sec:cost_period}
Figure~\ref{fig:cost_teaser} shows the cost of supporting 100 forward-in-time
analyses executed over different $\Delta t$s (x-axis) with a $50\%$ overlap.
\abbr{} is configured with a storage cache of size equal to the $25\%$ of the
total simulation data volume and a restart interval of $\Delta r = 8h$ 
%
The in-situ cost does not depend on $\Delta t$ since no data needs to be
permanently stored. On the contrary, the on-disk solution stores all the
simulation data, avoiding re-simulations. \abbr{} combines the two approaches:
while it requires less storage than on-disk, it needs to pay the cost of
re-simulating the missing files.
The cost-effectiveness of \abbr{} depends on the total
amount of analyses and $\Delta t$: if the data is analyzed by many
applications in a short availability period, then on-disk is a better because, once
the data is stored, the analysis is virtually free.  
Otherwise, if the same analyses are spread over a very long time
period, then in-situ is more cost-effective because no
(time-dependent) storage cost is paid.  
\abbr{} is designed to be cost-effective for scenarios
in between these two extremes: it does not store the full simulation data, saving
on the storage cost, but uses the storage to cache simulation data, saving
on the compute cost for recurrent analysis. 

Figure~\ref{fig:costmodel_time} shows this experiment varying the
\abbr{} cache size ($25\%$ and $50\%$) and $\Delta r$.  While larger
restart intervals require less storage for the restart files, they lead to an
increase of the \abbr{} cost for short $\Delta t$s: in these cases, the cost is
sensible to the re-simulations and larger $\Delta r$ can lead to more capacity
misses (it acts as cache block size, see Sec.~\ref{sec:virmod}).

\begin{figure}[h]
    \centering{
    \vspace{-1.1em}
    \includegraphics[trim={0 0 0 0}, clip, width=1\columnwidth]{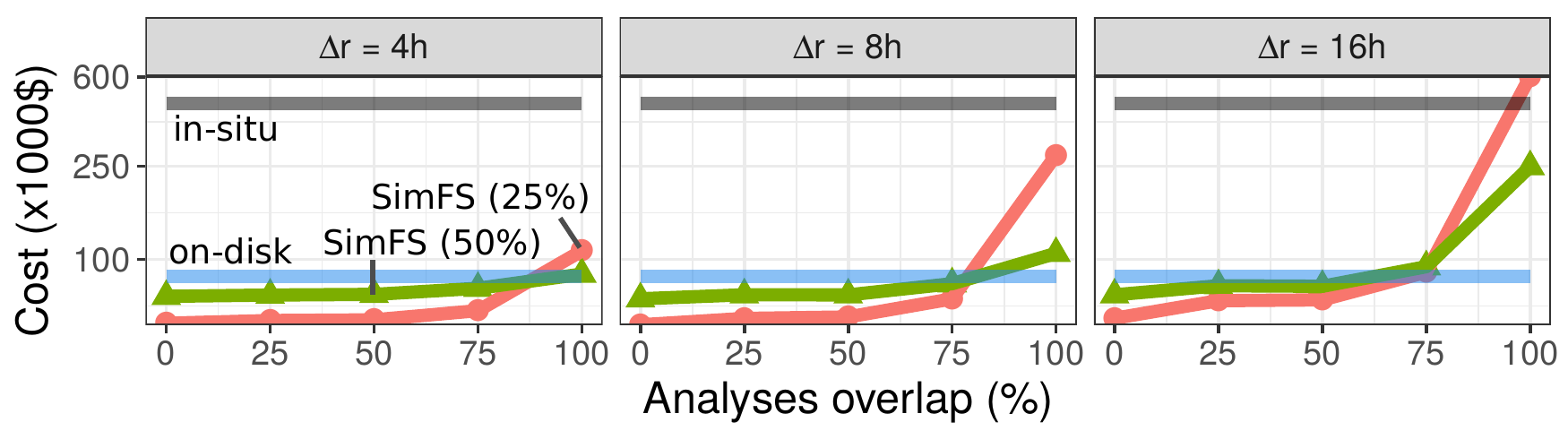}}
    \vspace{-2.0em}
    \caption{Data availability cost for different analyses overlaps.} 
    \label{fig:costmodel_overlap} 
    \vspace{-0.5em}
\end{figure} 


\textbf{Analyses Execution Overlap}
Figure~\ref{fig:costmodel_overlap} shows the same experiment but varying the
analyses overlaps and fixing \mbox{$\Delta t = 2y$} (other settings are
unchanged).  Higher overlap lead to more interleaved analyses: since they
access different output steps, this leads to a lower temporal locality,
hence to an increased number of misses.  This is amplified when using larger
$\Delta r$ since this can increase the number of capacity misses.

\begin{figure}[h]
    \centering{
    \vspace{-1.1em}
    \includegraphics[trim={0 0 0 0}, clip, width=1\columnwidth]{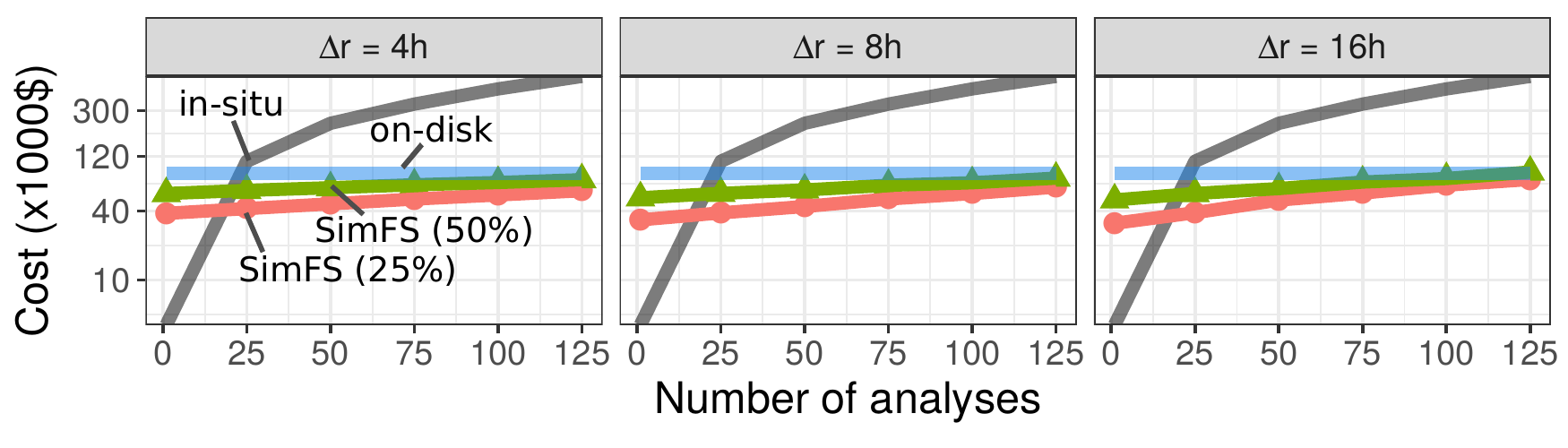}}
    \vspace{-2.0em}
    \caption{Data availability cost for different numbers of analyses.}
    \label{fig:costmodel_clients} 
    \vspace{-0.6em}
\end{figure} 

\enlargethispage*{\baselineskip}

\textbf{Total Number of Analyses}
Figure~\ref{fig:costmodel_clients} shows the cost when varying the number of
analyses executing during $\Delta t$. We fix the availability period to $\Delta
t = 2y$ and the analyses overlap at $50\%$.  
Independently from the restart interval and cache size, \abbr{} cannot beat
in-situ when the number of analyses is less than 20: the cost of the initial
simulation plus the storage of restarts and cached output steps is higher than
the cost of coupling each analysis with its own simulation. 
However, when increasing the number of analyses, in-situ becomes more
expensive since no data is shared among the different analyses.


\subsection{Discussion}


These cost models allow to estimate the data availability costs for both HPC
and cloud infrastructures: Figure~{\ref{fig:cost_discussion}a} is a heatmap showing the
ratio between the minimum cost between ondisk and in-situ and the \abbr{} cost,
for different storage and compute costs configurations (i.e., the darker the
color, the higher the ratio). 
We use the same scenario and parameters of Sec.~{\ref{sec:costeff}}, focusing
on the case with 100 analyses, 50\% overlap, $3y$ of data availability, and the
\abbr{} cache set up to the 25\% of the total simulation data volume.  
On the heatmap we show two real-world datapoints: the Microsoft Azure
configuration of Sec.~{\ref{sec:costeff}}, and the Piz Daint compute
and storage costs. The Piz Daint costs are derived from the CSCS cost
catalog~{\cite{cscsgo}}.

To determine the cost-effectiveness of \abbr{} w.r.t. other solutions, one
needs to know, among the others, the type of analyses performed during a given
data availability period. While this is a limiting factor of this cost model,
we plan to use online information to dynamically adapt the \abbr{}
configuration (e.g., cache size, restart interval) in a future work.
Figure~{\ref{fig:cost_discussion}b} and Figure~{\ref{fig:cost_discussion}c} show
the potential effects of these changes (same configuration of above). They
report the re-simulation cost and time as function of the storage space
reserved for restart files and for different cache sizes, respectively. They
show that (1) the restart interval and the cache size influence cost and
compute time, and (2) the reduced compute time due to having a bigger cache
might not be justified by the higher cost: e.g., for $\Delta r = 8$, a 50\%
cache size reduces the compute time of 20\% but increases the cost of 25\%.

\begin{center}
    \centering{
        \includegraphics[trim={0 0 0 0}, clip, width=1\linewidth ]{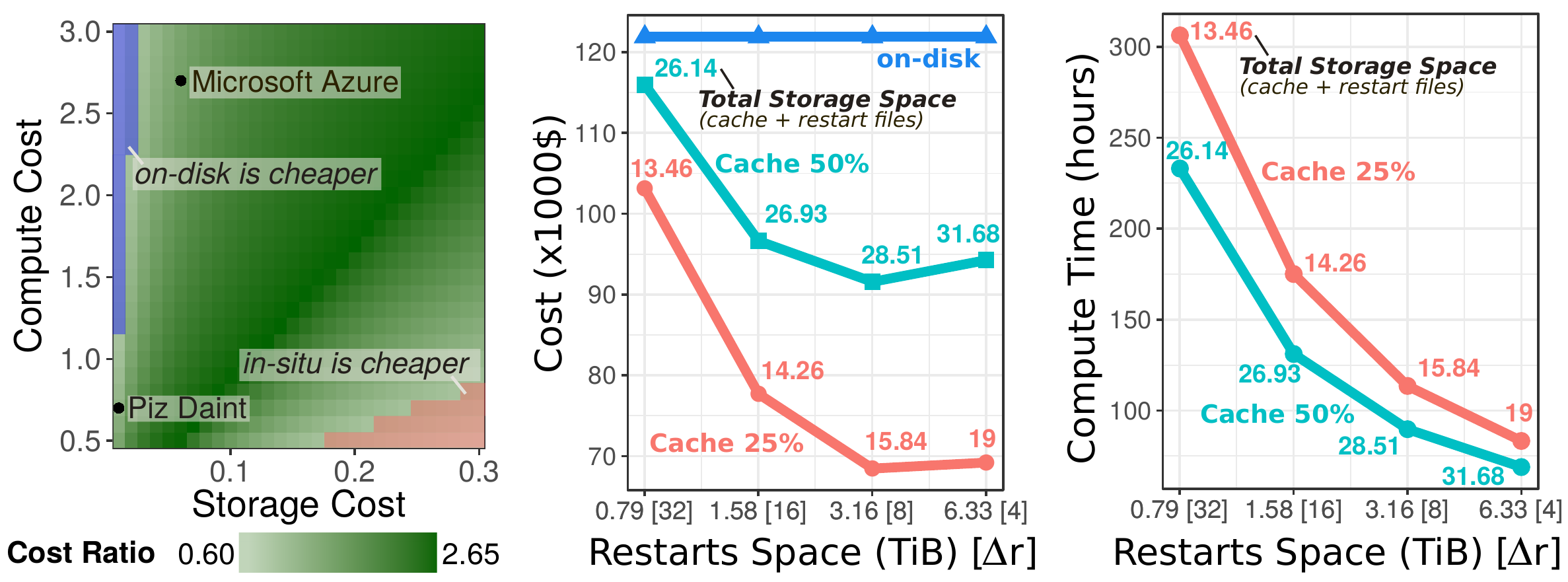}}
    \vspace{-1.8em}
    \captionof{figure}{(a) SimFS cost-effectiveness heatmap; (b) Cost over space; (c) Re-simulation time vs Space.}
    \label{fig:cost_discussion}
    \vspace{-0.7em}
\end{center}

\enlargethispage*{\baselineskip}

\section{Evaluation}\label{sec:evaluation}

The benchmarks presented in this section are executed on Piz Daint, a Cray XC50
system.  The compute nodes are equipped with two Intel Xeon E5-2695 @ 2.10GHz
with eighteen cores each. The system is interconnected with Cray's Aries
network and uses Lustre~\cite{braam2004lustre} as parallel file system.  The
measurements are taken by DVLib via the LibLSB library~\cite{benchmarking}.

\textbf{COSMO} is a non-hydrostatic local area atmospheric model used for both
operational numerical weather prediction and long-term climate
simulation~\cite{cosmo, fuhrer2017near}. 
In this benchmark we study the strong scalability of the system composed by a
virtualized COSMO simulation, \abbr{}, and a (sequential) analysis. 
The analysis computes mean and variance of a 1-D field of the simulation
output steps. The simulation proceeds in one-minute timesteps,
producing one output step every five minutes ($\Delta d = 5$) and one restart
file every hour ($\Delta r = 60$). 

The simulation context is configured to use the optimal number of compute nodes
($P=100$) as default, hence the prefetching strategy (2) (see
Sec.~\ref{sec:prefetching}) is applied.  Let us define $s_{\text{max}}$ as the
maximum number of re-simulations that can be run at the same time by
\abbr{}. This parameter limits the amount of computing resources that \abbr{}
can employ but it also limits the effectiveness of the prefetching strategies:
once $s_{\text{max}}$ simulations are running, \abbr{} will not be able to 
prefetch new ones to mask their restart latencies, delaying the analysis.

\begin{figure}[h]
    \vspace{-0.5em}
    \centering{ 
        \includegraphics[trim={10 6 0 0}, clip, width=0.9\columnwidth]{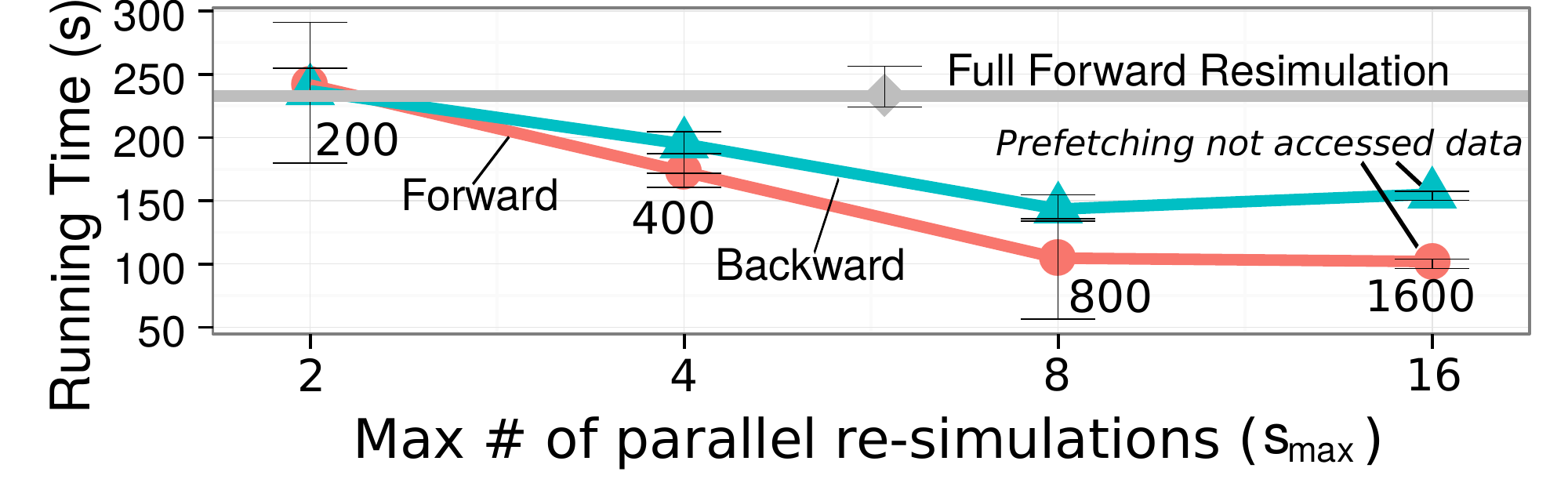}
    }
    \vspace{-0.5em}
    \caption{Strong scalability of analyses accessing
    virtualized COSMO data. The data points are annotated with the
    number of used compute nodes.}
    \label{fig:cosmo_scal} 
    \vspace{-1.6em}
\end{figure}
Figure~\ref{fig:cosmo_scal} shows the analysis completion time as function of
$s_{\text{max}}$. We report the completion times of a forward and backward
analysis accessing the same output steps but in different order. For
comparison, we also report the time of a full forward simulation, that is the
time needed by a single simulation to produce the same sequence of output
steps.
The analysis tool completion time scales up to a factor of $2.4$x w.r.t. the
full forward re-simulation when $s_{\text{max}}=8$. The backward simulation
shows a slightly worse scalability (up to a factor of $1.6$x): this is because
the first access of this analysis is served after the simulation of an entire
restart interval, delaying the prefetching activity (see
Figure~\ref{fig:pref_bw}).  
At $s_{\text{max}}=16$ prefetching does not bring any further benefit because
the prefetched simulations produce output steps that are not accessed by the
analysis, which terminates after analyzing the first $6$ hours of the simulated
data (i.e., $72$ output steps).


\begin{center}
    \centering{
    \vspace{-0.5em}
    \includegraphics[trim={0 5 0 0}, clip, width=1\columnwidth]{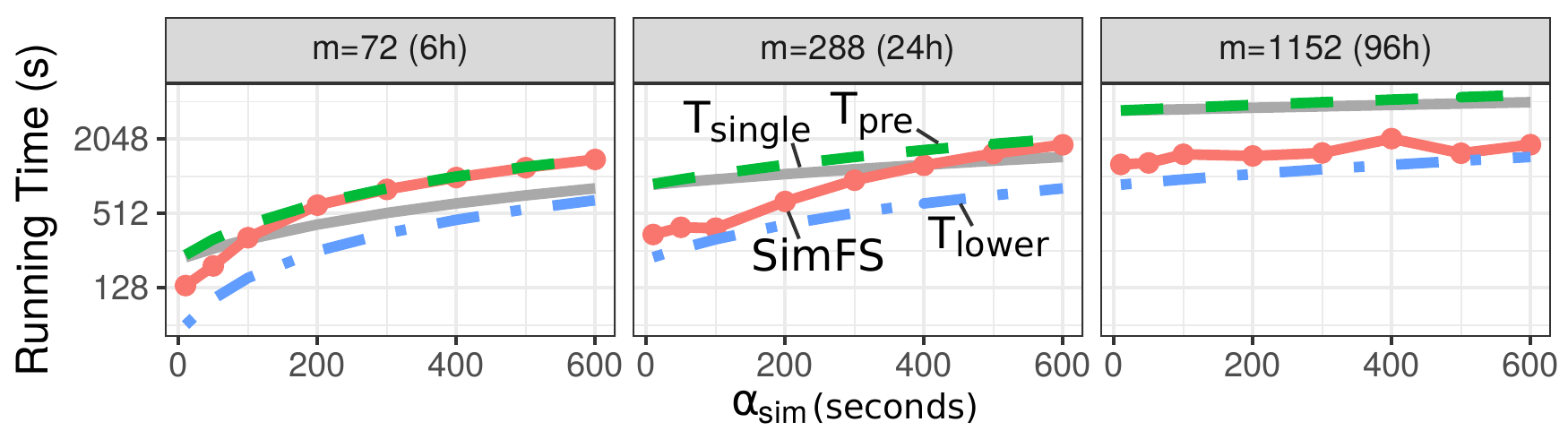}}
    \vspace{-1.6em}
    \captionof{figure}{Prefetching COSMO simulations under different restart latencies and analysis lengths.}
    \label{fig:cosmo_alpha} 
\end{center} 

With these settings, the simulation produces an output step every
\mbox{$\tau_{\text{sim}} = 3s$} on average and has restart latency of
\mbox{$\alpha_{\text{sim}} = 13s$}. The reported $\alpha_{\text{sim}}$ does not include the
re-simulation jobs queuing time. 
To study the effects the re-simulation jobs queuing times on the prefetching
effectiveness we simulate the analysis running time over different restart
latencies (now including the job queueing time) and analysis lengths ($m$).
We use a synthetic simulator that can be configured to
produce output steps at a given rate (i.e., $1/\tau_{\text{sim}}$) and after a given
restart latency. We use the same $\tau_{\text{sim}}$ of the
COSMO simulation described above, but we vary the restart latency in order to
simulate different job queuing times.
Figure~{\ref{fig:cosmo_alpha}} shows the results for $s_{\text{max}}=8$. 
As discussed in
Sec.~{\ref{sec:high_lat}}, when the restart latency is much higher than the
time needed to produce the output steps accessed by the analysis, the analysis
running time converges to the prefetching warm-up time and no benefits arise
from the prefetching of multiple simulations in parallel (i.e., strategy (2)). 
The warm-up time is a factor of two higher than
$T_{\text{single}}$, which is the time of a single simulation serving all the
analysis accesses: \mbox{$T_{\text{single}} = \alpha_{\text{sim}} + m \cdot \tau_{\text{sim}}$}.
This bounds the overhead that \abbr{} can introduce w.r.t. an in-situ analysis.
We also report a simple lower bound for this prefetching strategy,
$T_{\text{lower}}$, that is the given by the restart latency plus the time of
serving all the output steps requested by the analysis using $s_{\text{max}}$
simulations in parallel: \mbox{$T_{\text{lower}} = \alpha_{\text{sim}} + m \cdot
\frac{\tau_{\text{sim}}}{s_{\text{max}}}$}.

\enlargethispage*{2\baselineskip}

\textbf{FLASH} is a multiphysics simulation framework~\cite{fryxell2000flash}.
In this experiment we virtualize a Sedov simulation~\cite{sedov1959} which
involves the evolution of a blast wave from an initial pressure perturbation in
an otherwise homogeneous medium~\cite{flash44userguide}. The simulation is
configured to have $32^3$ cells per block (one block per core).  
We simulate the first second of the blast wave evolution. The simulation
proceeds in $0.005$s timesteps and produces one output step at each timestep
($\Delta d = 1$) and one restart file every $0.1$s ($\Delta r = 20$). 
The analysis computes mean and variance of the \textit{velocity} field. With
these settings, we measure $\tau_{\text{sim}} = 14s$ and $\alpha_{\text{sim}}=7s$ (not
including the re-simulation jobs queuing time).

\begin{figure}[h]
    \vspace{-0.5em}
    \centering{
        \includegraphics[trim={10 0 10 0}, clip, width=0.9\columnwidth]{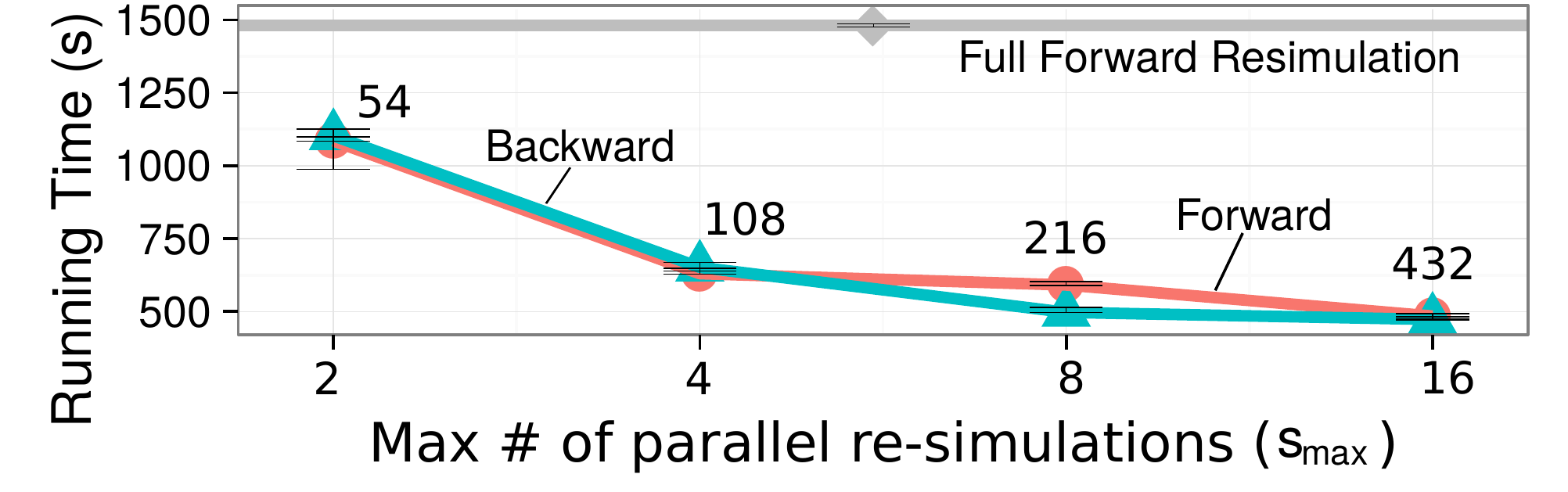}
    }
    \vspace{-0.8em}
    \caption{Strong scalability of analyses accessing  virtualized
    FLASH data. The data points are annotated with the
    number of used compute nodes.}
    \label{fig:flash_scal}
    \vspace{-0.5em}
\end{figure} 

Figure~\ref{fig:flash_scal} shows the analysis time over of $s_{\text{max}}$:
it scales up to a factor of $3$x when \mbox{$s_{\text{max}} = 16$}. 
Differently from the COSMO case, here forward and backward analysis show the
same behavior: This is due to the higher restart steps frequency of this
configuration that reduces the time needed to complete the resimulation
serving the first miss.

Figure~{\ref{fig:flash_alpha}} shows the analysis running time for different
restart latencies and analysis lengths, fixing $s_{\text{max}}=8$.  We
configure the synthetic simulator to run as the FLASH configuration described
above.
Differently from the COSMO study (i.e., Figure~{\ref{fig:cosmo_alpha}}), here the 
prefetching strategy is more effective: this is due to the number of output
steps analyzed and the higher $\tau_{\text{sim}}$, that better composate the
prefetching warm-up time $T_{\text{pre}}$.
This figure shows also how, in some cases, increasing the restart latency leads
to a reduction of the analysis running time (e.g., tile with $m=400$, between
$\alpha_{\text{sim}}=100s$ and $\alpha_{\text{sim}}=500s$). This is explained by the fact that,
due to the higher restart latency, \abbr{} determines a longer re-simulation
length $n$ (see Sec.~{\ref{sec:masklat}}), starting the simulation of the next
$s_{\text{max}} \cdot n$ output steps at each \textit{prefetch step}. The new
block of simulations may now simulate enough output steps to satisfy the remaining
analysis, avoiding the analysis  to pay a new restart latency caused by the
$s_{\text{max}}$ parameter.

\begin{center}
    \vspace{-0.5em}
    \includegraphics[trim={0 0 0 0}, clip, width=1\columnwidth]{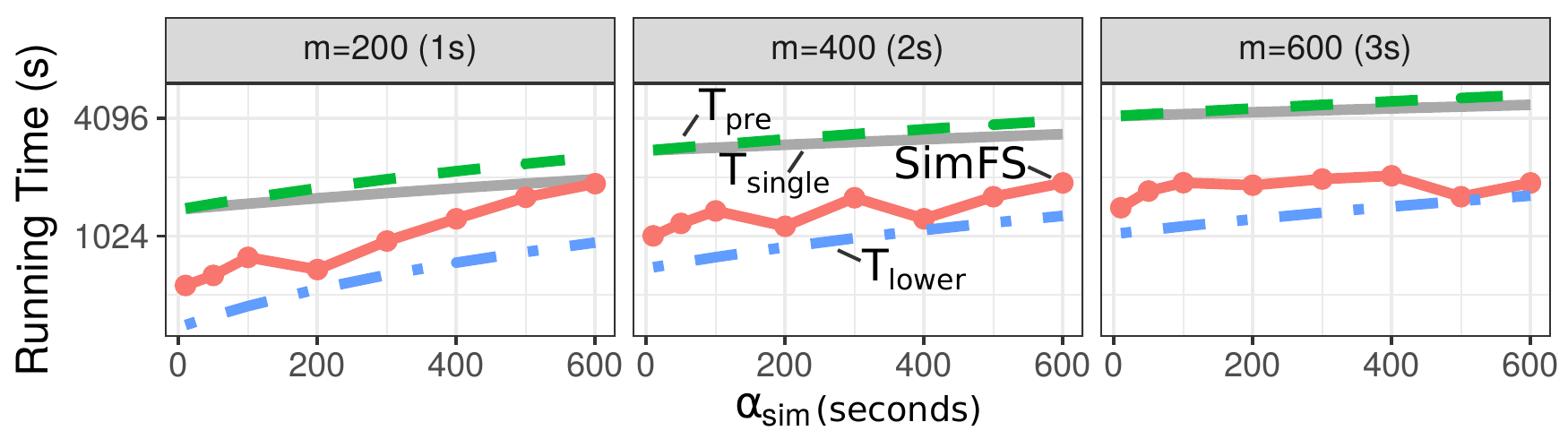}
    \vspace{-1.5em}
    \captionof{figure}{Prefetching FLASH simulations under different restart latencies and analysis lengths.}   
    \label{fig:flash_alpha} 
    \vspace{-0.5em}
\end{center}

\section{Related Work}

%
%

%
%

In-situ and on-disk are widely used solutions for simulation data analysis.
In-situ  avoids to store the data on disk by performing part of
the analysis (or filtering) directly at the simulation
site~\cite{richart2007toward, zhang2012enabling}, or during the data staging
phase (i.e., in transit)~\cite{bennett2012combining}, or at the analysis tool
site bypassing the parallel file system (i.e., loosely coupled
in-situ)~\cite{yu2010situ}. 
In all the cases, the analysis is performed as the data is simulated and
independent analysis applications have to run in tandem with their own
simulation.  
On-disk analysis is orthogonal to in-situ: here the analysis accesses the data
that is stored on disk, without dealing with the simulation process.
\abbr{} enables a tradeoff between these two approaches, which are at the two
ends of the storage requirements spectrum. In fact, virtualizing the simulation
output allows to adjust the storage requirements while offering to the
analysis the same file abstraction of the on-disk solution.

\abbr{} implements cache replacement strategies that are based on data locality
or data access cost.  Cost-based schemes are well studied in literature: Park
et al.~\cite{park2006cflru} consider different costs for writing back dirty
entries to flash memory disks and prioritize the eviction of the (cheaper)
non-dirty pages. However, this binary cost approach is not suitable in our
context where the output steps have costs linear in their distance from the
previous closest restart file. Jeong et al.~\cite{jeong2006cache,
jeong2003cost} propose a collection of cost-aware algorithms for NUMA
architectures with variable costs. Our cost-based replacement schemes build on
top of their algorithms (i.e., BCL and DCL).

\section{Summary}

We argue that storing the full simulation output is not cost-effective
because the ever growing availability of computing power enables multi-petabyte
simulation runs.
\abbr{} virtualizes the simulation data: the data is only partially stored
and accesses to missing data are served by restarting simulations. 
The analysis applications can be transparently interfaced to \abbr{} or made
virtualization-aware by using the \abbr{} APIs.

All in all, \abbr{} introduces a new simulation data analysis paradigm that
relaxes the storage requirements and offers a viable path towards exa-scale
simulations.
\abbr{} can be downloaded at:

\begin{center}
\texttt{https://github.com/spcl/SimFS}
\end{center}

\section*{Acknowledgments} 
This work was supported by the Swiss National Science Foundation under Sinergia
grant CRSII2\_154486/1 and by a grant from the Swiss National Supercomputing
Centre and PRACE.  We acknowledge ECMWF for providing the access
traces dataset. We thank Christoph Sch\"{a}r, David Leutwyler, and all the members
of the crCLIM project for the great discussions.
%

 
\enlargethispage*{2\baselineskip}

\vspace{-1em}
{\scriptsize
\setlength{\parskip}{0pt}
\setlength{\bibsep}{0pt plus 1ex}
\bibliographystyle{IEEEtran}
\textsf{\bibliography{bibliography}}}

\begin{thebibliography}{10}
\providecommand{\url}[1]{#1}
\csname url@samestyle\endcsname
\providecommand{\newblock}{\relax}
\providecommand{\bibinfo}[2]{#2}
\providecommand{\BIBentrySTDinterwordspacing}{\spaceskip=0pt\relax}
\providecommand{\BIBentryALTinterwordstretchfactor}{4}
\providecommand{\BIBentryALTinterwordspacing}{\spaceskip=\fontdimen2\font plus
\BIBentryALTinterwordstretchfactor\fontdimen3\font minus
  \fontdimen4\font\relax}
\providecommand{\BIBforeignlanguage}[2]{{%
\expandafter\ifx\csname l@#1\endcsname\relax
\typeout{** WARNING: IEEEtran.bst: No hyphenation pattern has been}%
\typeout{** loaded for the language `#1'. Using the pattern for}%
\typeout{** the default language instead.}%
\else
\language=\csname l@#1\endcsname
\fi
#2}}
\providecommand{\BIBdecl}{\relax}
\BIBdecl

\bibitem{Grawinkel:2015:AES:2750482.2750484}
\BIBentryALTinterwordspacing
M.~Grawinkel \emph{et~al.}, ``{Analysis of the ECMWF Storage Landscape},'' in
  \emph{Proc. of the 13th USENIX Conf. on File and Storage Technologies}, ser.
  FAST'15.\hskip 1em plus 0.5em minus 0.4em\relax Berkeley, CA, USA: USENIX
  Association, 2015. [Online]. Available:
  \url{http://dl.acm.org/citation.cfm?id=2750482.2750484}
\BIBentrySTDinterwordspacing

\bibitem{potter2016pkdgrav3}
D.~Potter \emph{et~al.}, ``{PKDGRAV3: Beyond Trillion Particle Cosmological
  Simulations for the Next Era of Galaxy Surveys},'' \emph{arXiv preprint
  arXiv:1609.08621}, 2016.

\bibitem{bernyk2016theoretical}
M.~Bernyk \emph{et~al.}, ``The theoretical astrophysical observatory:
  Cloud-based mock galaxy catalogs,'' \emph{The Astr. Journal Supplement
  Series}, vol. 223, no.~1, p.~9, 2016.

\bibitem{otv}
A.~Ragagnin \emph{et~al.}, ``An online theoretical virtual observatory for
  hydrodynamical, cosmological simulations,'' \emph{arXiv preprint
  arXiv:1612.06380}, 2016.

\bibitem{winsberg2010science}
E.~Winsberg, \emph{Science in the age of computer simulation}.\hskip 1em plus
  0.5em minus 0.4em\relax University of Chicago Press, 2010.

\bibitem{folk1999hdf5}
M.~Folk \emph{et~al.}, ``{HDF5: A file format and I/O library for high
  performance computing applications},'' in \emph{Proc. of Supercomputing},
  vol.~99, 1999.

\bibitem{rew1990unidata}
R.~K. Rew and G.~P. Davis, ``{The unidata netCDF: Software for scientific data
  access},'' in \emph{Sixth Int. Conf. on Interactive Information and
  Processing Systems for Meteorology, Oceanography, and Hydrology}, 1990.

\bibitem{lofstead2008flexible}
J.~F. Lofstead \emph{et~al.}, ``{Flexible IO and integration for scientific
  codes through the adaptable IO system (ADIOS)},'' in \emph{Proc. of the 6th
  Int. Workshop on Challenges of Large Applications in Distributed
  Environments}.\hskip 1em plus 0.5em minus 0.4em\relax ACM, 2008.

\bibitem{arteaga-bitrep}
A.~Arteaga, O.~Fuhrer, and T.~Hoefler, ``{Designing Bit-Reproducible Portable
  High-Performance Applications},'' in \emph{Proc. of the 28th IEEE Int.
  Parallel and Distributed Processing Symp. (IPDPS)}.\hskip 1em plus 0.5em
  minus 0.4em\relax IEEE Computer Society, Apr. 2014.

\bibitem{muller2018reproducible}
I.~M{\"u}ller \emph{et~al.}, ``{Reproducible Floating-Point Aggregation in
  RDBMSs},'' \emph{arXiv preprint arXiv:1802.09883}, 2018.

\bibitem{Hill:2014:CCA:2692916.2558890}
\BIBentryALTinterwordspacing
M.~D. Hill, ``{21st Century Computer Architecture},'' \emph{SIGPLAN Not.}, Feb.
  2014. [Online]. Available: \url{http://doi.acm.org/10.1145/2692916.2558890}
\BIBentrySTDinterwordspacing

\bibitem{jiang2002lirs}
S.~Jiang and X.~Zhang, ``{LIRS: an efficient low inter-reference recency set
  replacement policy to improve buffer cache performance},'' \emph{ACM
  SIGMETRICS Performance Evaluation Review}, 2002.

\bibitem{megiddo2003arc}
N.~Megiddo and D.~S. Modha, ``{ARC: A Self-Tuning, Low Overhead Replacement
  Cache},'' in \emph{FAST}, 2003.

\bibitem{jeong2006cache}
J.~Jeong and M.~Dubois, ``Cache replacement algorithms with nonuniform miss
  costs,'' \emph{IEEE Transactions on Computers}, vol.~55, no.~4, pp. 353--365,
  2006.

\bibitem{azure}
``{Microsoft Azure},'' \url{https://azure.microsoft.com}, 2018, accessed:
  2018/04.

\bibitem{azure-vm-pricing}
``{Microsoft Azure VM Pricing},''
  \url{https://azure.microsoft.com/en-us/pricing/details/virtual-machines/linux/},
  2018, accessed: 2018/04.

\bibitem{azure-storage-pricing}
``{Microsoft Azure Storage Pricing},''
  \url{https://azure.microsoft.com/en-us/pricing/details/storage/files/}, 2018,
  accessed: 2018/04.

\bibitem{cscsgo}
``{CSCS2Go},'' \url{https://2go.cscs.ch/home/}, 2018, accessed: 2018/12.

\bibitem{braam2004lustre}
P.~J. Braam \emph{et~al.}, ``{The Lustre storage architecture},'' 2004.

\bibitem{benchmarking}
T.~Hoefler and R.~Belli, ``{Scientific Benchmarking of Parallel Computing
  Systems}.''\hskip 1em plus 0.5em minus 0.4em\relax ACM, Nov. 2015, pp.
  73:1--73:12, proc. of the Int. Conf. for High Performance Computing,
  Networking, Storage and Analysis (SC15).

\bibitem{cosmo}
\BIBentryALTinterwordspacing
{Consortium for small-scale modeling}, ``Web page,'' 1998, accessed: 2017/03.
  [Online]. Available: \url{http://www.cosmo-model.org}
\BIBentrySTDinterwordspacing

\bibitem{fuhrer2017near}
O.~Fuhrer \emph{et~al.}, ``Near-global climate simulation at 1km resolution:
  establishing a performance baseline on 4888 gpus with cosmo 5.0,''
  \emph{Geoscientific Model Development Discussions}, 2017.

\bibitem{fryxell2000flash}
\BIBentryALTinterwordspacing
B.~Fryxell \emph{et~al.}, ``{FLASH: An Adaptive Mesh Hydrodynamics Code for
  Modeling Astrophysical Thermonuclear Flashes},'' \emph{The Astr. Journal
  Supplement Series}, vol. 131, no.~1, p. 273, 2000. [Online]. Available:
  \url{http://stacks.iop.org/0067-0049/131/i=1/a=273}
\BIBentrySTDinterwordspacing

\bibitem{sedov1959}
L.~I. {Sedov}, \emph{{Similarity and Dimensional Methods in Mechanics}}, 1959.

\bibitem{flash44userguide}
\BIBentryALTinterwordspacing
F.~C. for Computational Science University~of Chicago, ``Flash user's guide
  version 4.4,'' 2016, accessed: 2017-03-17. [Online]. Available:
  \url{http://flash.uchicago.edu/site/flashcode/user_support/flash4_ug_4p4.pdf}
\BIBentrySTDinterwordspacing

\bibitem{richart2007toward}
N.~Richart \emph{et~al.}, ``Toward a computational steering environment for
  legacy coupled simulations,'' in \emph{Sixth Int. Symp. on Par. and Distr.
  Comp., 2007. (ISPDC'07)}.\hskip 1em plus 0.5em minus 0.4em\relax IEEE, 2007.

\bibitem{zhang2012enabling}
F.~Zhang \emph{et~al.}, ``Enabling in-situ execution of coupled scientific
  workflow on multi-core platform,'' in \emph{26th Int. Par. \& Distr.
  Processing Symp. (IPDPS)}.\hskip 1em plus 0.5em minus 0.4em\relax IEEE, 2012.

\bibitem{bennett2012combining}
J.~C. Bennett \emph{et~al.}, ``Combining in-situ and in-transit processing to
  enable extreme-scale scientific analysis,'' in \emph{Int. Conf. for High
  Performance Computing, Networking, Storage and Analysis (SC), 2012}.\hskip
  1em plus 0.5em minus 0.4em\relax IEEE, 2012, pp. 1--9.

\bibitem{yu2010situ}
H.~Yu \emph{et~al.}, ``In situ visualization for large-scale combustion
  simulations,'' \emph{IEEE computer graphics and applications}, vol.~30,
  no.~3, 2010.

\bibitem{park2006cflru}
S.-y. Park \emph{et~al.}, ``{CFLRU: a replacement algorithm for flash
  memory},'' in \emph{Proc. of the 2006 Int. Conf. on Compilers, Architecture
  and Synthesis for Embedded Systems}.\hskip 1em plus 0.5em minus 0.4em\relax
  ACM, 2006.

\bibitem{jeong2003cost}
J.~Jeong and M.~Dubois, ``Cost-sensitive cache replacement algorithms,'' in
  \emph{Proc. The Ninth Int. Symp. on High Performance Architecture}.\hskip 1em
  plus 0.5em minus 0.4em\relax IEEE, 2003.

\end{thebibliography}

\end{document}